# Localized High-Concentration Electrolytes Get More Localized Through Micelle-Like Structures

Corey M. Efaw,[1,2,#] Qisheng Wu,[3,#] Ningshengjie Gao,[1] Yugang Zhang,[4] Haoyu Zhou,[2] Kevin Gering,[1] Michael F. Hurley,[2] Hui Xiong,[2] Enyuan Hu,[5] Xia Cao,[6] Wu Xu,[6] Ji-Guang Zhang,[6] Eric J. Dufek,[1] Jie Xiao,[6,7] Xiao-Qing Yang,[5] Jun Liu,[6,7] Yue Qi,[3,*] and Bin Li[1,2,*]

[1] Energy and Environmental Science and Technology, Idaho National Laboratory, Idaho Falls, ID 83415, USA

[2] Micron School of Materials Science and Engineering, Boise State University, Boise, ID 83725, USA

[3] School of Engineering, Brown University, Providence, RI 02912, USA

[4] Center for Functional Nanomaterials, Brookhaven National Laboratory, Upton, NY 11973, USA

[5] Chemistry Division, Brookhaven National Laboratory, Upton, NY 11973, USA

[6] Energy and Environment Directorate, Pacific Northwest National Laboratory, Richland, WA, 99252 USA

[7] Materials Science and Engineering Department, University of Washington, Seattle, WA, 98105 USA

#These authors contributed equally to this work

*Corresponding Authors: bin.li@inl.gov and yueqi@brown.edu

# Abstract

Liquid electrolytes in batteries are typically treated as macroscopically homogeneous ionic transport media despite having complex chemical composition and atomistic solvation structures, leaving a knowledge gap of microstructural characteristics. Here, we reveal a unique micelle-like structure in a localized high-concentration electrolyte (LHCE), in which the solvent acts as a surfactant between an insoluble salt in diluent. The miscibility of the solvent with the diluent and simultaneous solubility of the salt results in a micelle-like structure with a smeared interface and an increased salt concentration at the centre of the salt-solvent clusters that extends the salt solubility. These intermingling miscibility effects have temperature dependencies, wherein an exemplified LHCE peaks in localized cluster salt concentration near room temperature and is utilized to form a stable solid-electrolyte interphase (SEI) on Li-metal anode. These findings serve as a guide to predicting a stable ternary phase diagram and connecting the electrolyte microstructure with electrolyte formulation and formation protocols to form stable SEI for enhanced battery cyclability.

Liquid electrolytes play a critical role in developing high-energy rechargeable batteries needed to advance electric vehicle capabilities. Conventional low-concentration electrolytes (LCEs) need to be replaced to make long-life batteries a reality. The solvent-derived, instable, and heterogeneous solid-electrolyte interphase (SEI) layers formed on high-capacity anodes, such as lithium (Li) metal, silicon (Si), sodium (Na) metal, zinc (Zn) metal and black phosphorus (BP), cannot accommodate large volume changes, leading to continuous loss of active materials and rapid dendrite growth.

One of the key pathways to harnessing highly reactive, yet energetic anodes is by regulating electrolyte solvation structures beyond that of LCEs.[1,2] Increasing the salt concentration to form high-concentration electrolytes (HCEs) enables preferential anion reduction to form stable, inorganic-rich SEI and reduce parasitic reactions of free solvent molecules.[3-7] However, increasing salt concentration results in sluggish ion transport .[5] To mitigate these pitfalls, a low viscosity diluent is added to form localized high-concentration electrolytes (LHCEs), thus improving high-capacity anode performance (e.g., Li,[8-11] Si,[12,13] Na,[14] Zn,[15,16] and BP[17,18]).

Previous LHCE experiments and computations show that the cation solvation shells are fully occupied by the salt anion and solvent with minimal diluent participation.[9,19] Salt-solvent clusters of ~1 nm[20] are believed to retain a random, relatively uniform distribution (**Fig. 1a**), much like those in HCEs.[7] However, the information about LHCE microstructures, which bridges the scales from atomistic solvation structures to macroscopically homogeneous liquid electrolyte, is still missing, leaving many unanswered questions. For example, why don't diluent molecules participate in the solvation shell?[21,22]  Do the salt-solvent clusters agglomerate uniformly? Why does LHCE improve performance versus HCE of the same salt-to-solvent molar ratio? In this paper, we propose a micelle-like structure in LHCE to unify the answers to these questions, by combining molecular dynamics (MD) simulations and Raman spectroscopy, along with small-angle/wide-angle X-ray scattering (SAXS-WAXS) for validation.

Analogical to the micelle concept for dispersed emulsions of non-mixing substances, [23,24] the salt is insoluble in the diluent in an LHCE, while the diluent is miscible with the solvent.[7] A ternary phase diagram illustrates the interactions between salt, solvent, and diluent and further demonstrates that the solvent acts as a surfactant, binding immiscible salt and diluent phases, reducing the interfacial energy, and stabilizing the dispersed liquid microstructure. The newly proposed micelle-like structure in LHCE (**Fig. 1b**) is based on the simulated structures of lithium bis(fluorosulfonyl)imide (LiFSI) salt, dimethoxyethane (DME) solvent, and tris(2,2,2-trifluoroethyl)orthoformate (TFEO) diluent in coordination with analytical and electrochemical measurements. The solvent differs from traditional surfactant molecules, which typically have a polar-philic head and polar-phobic tail (e.g., hydrophilic/phobic water/oil emulsion[24] or lithiophilic/phobic hydrofluoroether-based electrolyte, **Fig. 1c**).[25] For this reason, LHCE is referred to as "micelle-like", where a network of salt-solvent clusters are mostly separated from the diluent matrix by a solvent-rich surfactant region (**Fig. 1b**). While this micelle-like structure is consistent with the previously proposed solvation structures of LHCEs,[7] it further explains why those clusters are stable and improve upon their HCE counterparts.

The distribution of DME molecules is a result of minimizing the free energy of the ternary system. In addition to the interface region, a small fraction exists in the miscible diluent matrix (**Fig. 1b**). This explains the observation of increased free solvent molecules with increasing diluent concentration.[26] DME in the matrix or near the interface region will naturally increase salt aggregation at the centre of the salt-solvent network. **Fig. 1b** is idealized as a refined, circular network of clusters, while in reality it can be more complex in its shape and size.[27]

Moreover, both salt-solvent and solvent-diluent interactions existing in the micelle-like structure are temperature-sensitive, changing DME solvent distribution and local salt concentration. While many temperature- and rate-dependent SEI formation protocols have been proposed,[28] detailed mechanisms are unclear. Here, we demonstrate that the salt-solvent and solvent-diluent interactions impose different temperature dependencies. This is exemplified with a LiFSI-1.2DME-2TFEO LHCE by mol, where a local salt concentration peak at 25°C is observed within the 10-45°C temperature range, resulting in improved SEI composition and morphology, along with cycling performance. An additional LHCE formulation of LiFSI-dimethyl carbonate (DMC)-TTE was formulated to balance an improved micelle-like cluster network with macroscopic properties, resulting in an improved Coulombic efficiency (CE) when compared to literature. These findings suggest that controlling the underlying microstructure of LHCE, through optimization of the electrolyte component contributions and external parameters, directly impacts SEI design and battery optimization.

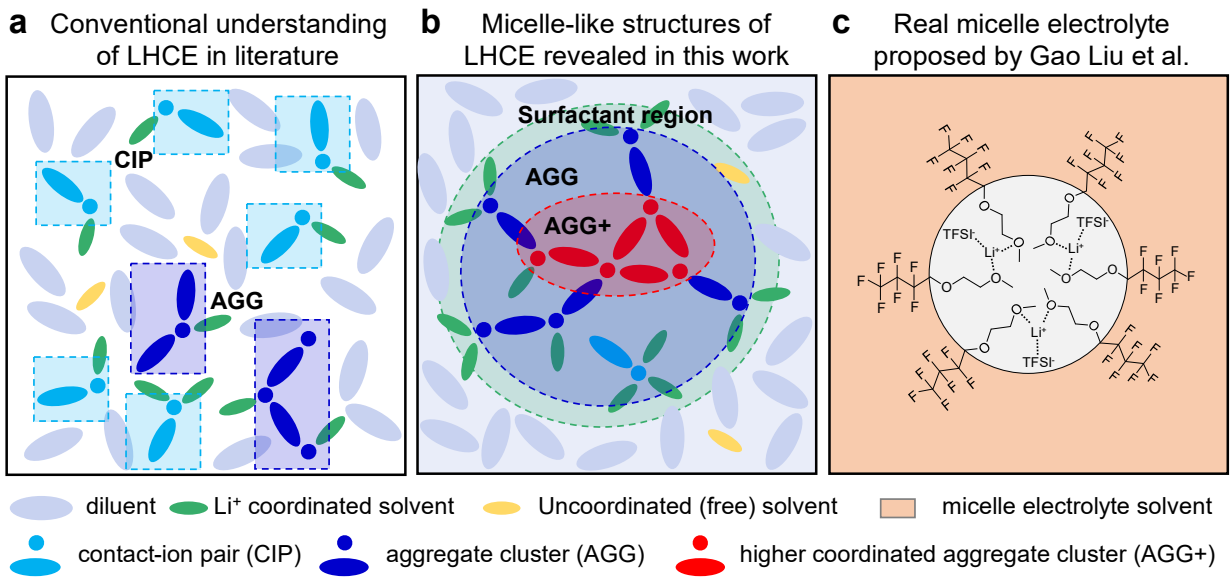

**Fig. 1 | Schematics for the conventional understanding of LHCE, micelle-like structure of LHCE and the real micelle electrolyte. a**, Schematic for the conventional understanding of LHCE in literature.[7] The light blue and purple areas refer to diluent and high-concentration salt-solvent clusters, respectively, where the clusters are maintained as they are in HCE. **b**, Schematic for the micelle-like structure of LHCE revealed in this work (AGG: ion-pair aggregates; AGG+: more coordinated ion-pair aggregates; CIP: contact ion pairs). **c**, The real micelle electrolyte (lithiophilic/phobic hydrofluoroether-based solvent in a lithium bis(trifluoromethanesulfonyl)imide (LiTFSI)/1,1,2,2-tetrafluoroethyl-2,2,3,3-tetrafluoropropyl ether (TTE) electrolyte) proposed by Gao Liu et al, which is reproduced here from their work.[25]

**Micelle-Like Structure Characteristics in LHCE**

A ternary phase diagram and MD-simulated atomic structures of mixed LiFSI salt, DME solvent, and TFEO diluent, are provided in **Fig. 2a**. First, DME dissolves LiFSI up to a solubility limit (~1:1.05 LiFSI:DME by mol). Simulated HCEs (LiFSI-1.2DME and LiFSI-1.4DME) and LCE (LiFSI-9DME) are shown in **Fig. 2b-d** and Supplementary Fig. 1a-c. With Raman Spectroscopy, the C-O stretching vibration mode of pure DME peaks (820-850 cm$^{-1}$) are reduced after LiFSI is dissolved in DME, blue-shifting to 873-877 cm$^{-1}$,[29] corresponding to Li$^+$ binding to ether oxygen atoms (**Fig. 3a**), as confirmed with MD simulations (Supplementary Fig. 2). DME and TFEO are miscible (**Fig. 2e**), while LiFSI has minimal or no solubility in TFEO, as confirmed by MD (**Fig. 2f**) and Raman analysis, where peaks are retained in the 820-870 cm$^{-1}$ range between TFEO and LHCE (**Fig. 3a**). Combining these component interactions reveal the solvent as a surfactant in LHCE, where TFEO has almost-zero contribution to the Li$^+$ solvation shell (Supplementary Fig. 1d-e), while DME exists mostly within the network of salt-solvent clusters with few in the TFEO matrix. The size and shape of the network composed of salt-solvent clusters can vary when the concentration of salt, solvent, and diluent changes (**Fig. 2g-h**).

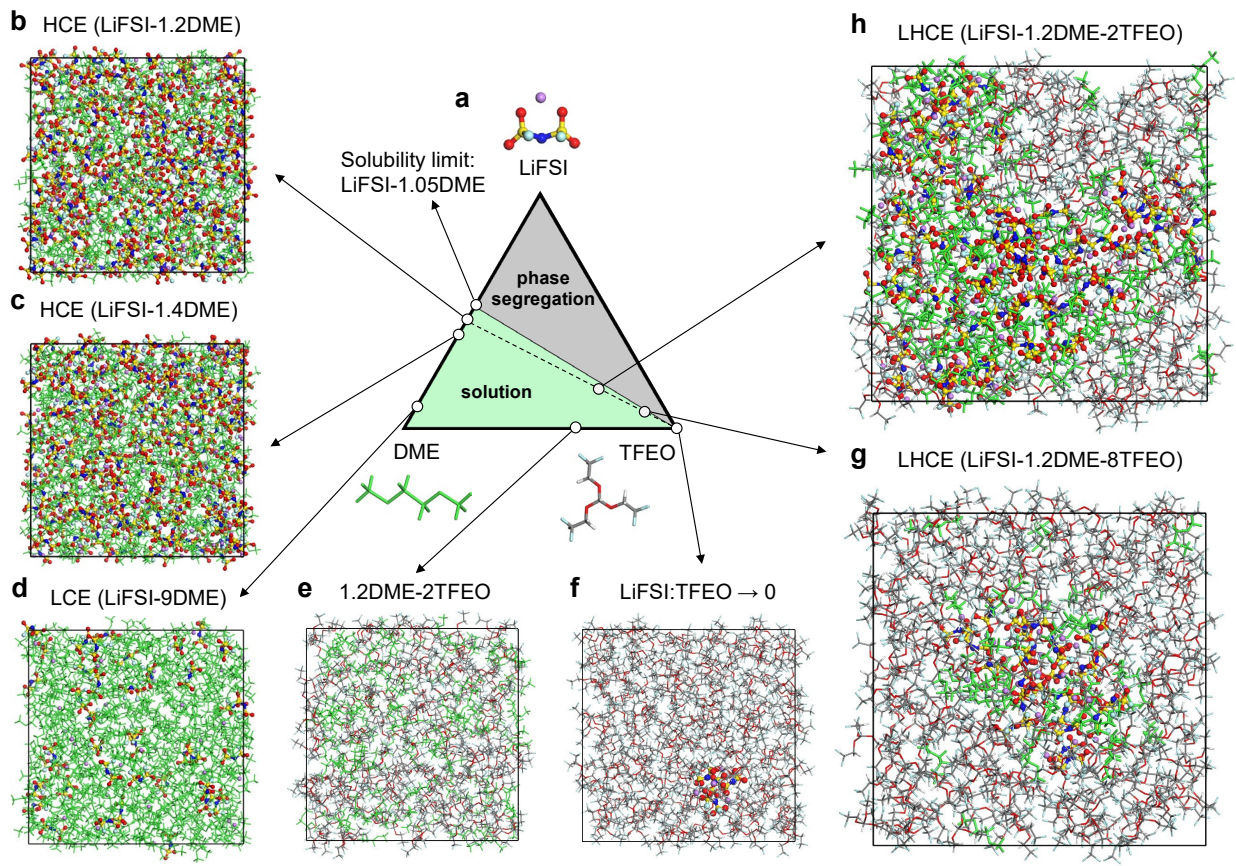

**Fig. 2 | Ternary phase diagram of LiFSI salt, DME solvent, and TFEO diluent**. **a**, Ternary phase diagram connecting the three variable phases: LiFSI, DME, and TFEO. The coloured ball-and-stick model stands for LiFSI, the green stick model for DME, and the coloured stick model for TFEO. The ternary phase diagram is divided into two regions, the solution phase (green area) and the phase segregation phase (i.e., insoluble salt, grey area). MD-simulated structures of **b**, HCE (LiFSI-1.2DME) and **c**, HCE (LiFSI-1.4DME), showing uniformly distributed $Li^+$-$FSI^-$ clusters. The HCE (LiFSI-1.05DME) near the solubility limit is also noted on the phase diagram. **d**, MD-simulated structure of LCE (LiFSI-9DME), showing uniformly distributed $Li^+$-$FSI^-$ clusters. **e**, MD-simulated structure of the mixed solvent and diluent (1.2DME-2TFEO), revealing high miscibility between DME solvent and TFEO diluent. **f**, MD-simulated structure of 4 LiFSI molecules in TFEO matrix, revealing no solvation of LiFSI salt in TFEO. The cations and anions were initially and uniformly separated in the TFEO diluent and formed the small cluster of 4 $Li^+$ and 4 $FSI^-$ by the end of the simulation. MD-simulated structures of **g**, LHCE (LiFSI-1.2DME-8TFEO) and **h**, LHCE (LiFSI-1.2DME-2TFEO), in both of which the network of salt-solvent clusters is surrounded by TFEO diluent matrix.

The ability of solvents and diluents to solvate Li$^+$ is thought to be reflected by dielectric constant and donor number,[21] but subject to debate. DME and TFEO have a similar dielectric constant (~7.0)[22,30] and similar binding energies to Li$^+$ for single molecules (2.81 eV *vs* 2.00 eV, Supplementary Fig. 3a-b). However, to form a solvation shell, the number of solvating molecules depends on the geometry of and the interaction with the solvated atom/molecule (e.g., four to five ethylene carbonate (EC) molecules,[31] three DME molecules, or two TFEO molecules). When Li$^+$ coordinates with three DME molecules, the binding energy is comparable to a Li$^+$-FSI$^-$ ion pair (5.39 eV versus 6.07 eV, Supplementary Fig. 3c-d). Li$^+$ coordinating to two TFEO molecules exhibits a lower binding energy of 2.89 eV (Supplementary Fig. 3e), driven by steric and electronic effects. Thus, despite their comparable dielectric constants, DME solvates LiFSI (**Fig. 2b-d**) while TFEO does not (**Fig. 2f**). As solubility reflects the interactions between the binary systems, it serves as a stronger descriptor than dielectric constants for LHCEs. Ultimately, the formation of the micelle-like structure is maintained through a competition of energy of mixing and interfacial interactions. The competition of these interactions computed at quantum level is carried into MD simulations to ensure the accuracy of the liquid structures (Methods).

**Higher Local Salt Concentration Through Micelle-like Structures in LHCE**

The micelle-like structure pushes the local salt concentration in LHCE higher than its HCE counterpart, which is validated by Raman spectroscopy. The Li$^+$-FSI$^-$ coordination strength is characterized by the S-N-S symmetric stretching vibrational mode (715-780 cm$^{-1}$, **Fig. 3a**).[9] Solid LiFSI salt (~775 cm$^{-1}$)[32] red-shifts when dissolved in DME solvent, driven by high sensitivity to Li$^+$-FSI$^-$ Coulombic interactions.[33] The peak further red-shifts as salt concentration decreases, going from ~753 cm$^{-1}$ near the solubility limit (LiFSI-1.05DME) to ~749 cm$^{-1}$ for LiFSI-1.2DME, ~746 cm$^{-1}$ for LiFSI-1.4DME, and ~721 cm$^{-1}$ for LiFSI-9DME. Comparably, LHCE (LiFSI-1.2DME-2TFEO) peaks at ~752 cm$^{-1}$, blue-shifting from HCE with the same salt-to-solvent molar ratio (LiFSI-1.2DME), suggesting higher local salt concentration in LHCE. Furthermore, Raman deconvolution analysis quantified contributions of cluster interactions. Following literature,[4,5,10] solvent-separated ion pairs (SSIP), contact ion pairs (CIP), ion-pair aggregates (AGG), and more coordinated ion-pair aggregates (AGG+) were defined by an increase in anion-cation association (**Fig. 3b**).[33] LCE is dominated by SSIP and CIP, while HCEs and LHCE are prominently AGG and AGG+. Notably, the ratio of AGG+ in LHCE (51.4%) is higher than that its HCE counterpart (40.4%), indicating stronger $Li^+$-$FSI^-$ association.

In parallel, MD simulations and coordination analyses were conducted. The salt-solvent clusters (SSIP, CIP, AGG, and AGG+) are categorized based on the $FSI^-$-$Li^+$ coordination number (CN, Supplementary Fig. 4). As shown in **Fig. 2h** and **Fig. 3c**, LiFSI and DME form a three-dimensional network of connected salt-solvent clusters surrounded by a TFEO matrix. A salt concentration gradient is exhibited within the salt-solvent clusters, where AGG+ trends to stay at the centre of the network, while AGG resides nearer the outer shell (**Fig. 3d**), differing from homogeneous spatial distributions of clusters in binary electrolytes (Supplementary Fig. 5). Additionally, DME-$Li^+$ interactions accumulate at the cluster-network/matrix interface, playing the role of surfactant (**Fig. 3c-e**). Furthermore, a fraction of free DME molecules is completely dissolved in the miscible TFEO matrix (**Fig. 3c, e**), which further enhances the local salt concentration.

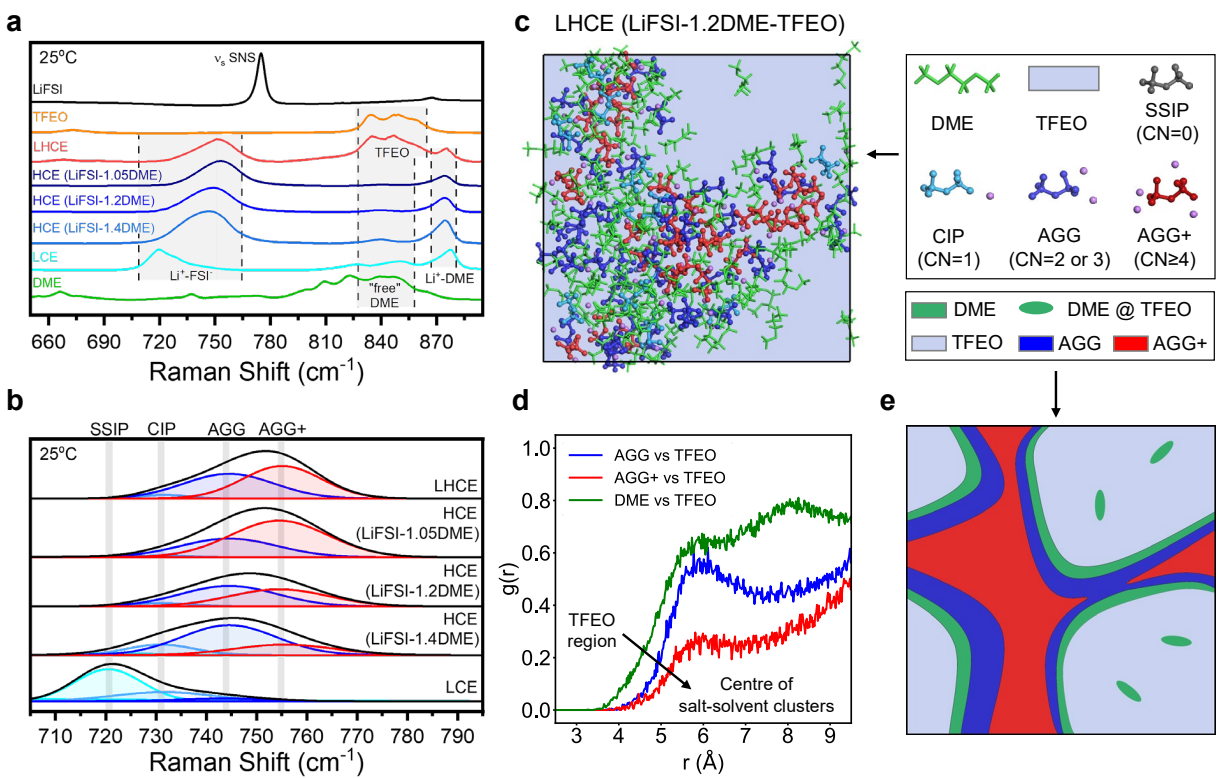

**Fig. 3 | Raman spectroscopy and MD simulations of different systems at 25 °C. a**, Raman spectra at 25°C for (top to bottom) solid LiFSI crystal (black), TFEO (orange), LHCE (LiFSI-1.2DME-2TFEO, red), various HCEs (LiFSI-1.05DME as navy, LiFSI-1.2DME as blue, and

LiFSI-1.4DME as light blue), LCE (LiFSI-9DME, cyan), and DME (green). Different peak ranges are noted in the spectra. **b**, Deconvolution of Li$^+$-FSI$^-$ Raman peaks for LHCE, HCEs, and LCE with peak fits for different cluster types denoted. **c**, MD trajectory snapshot showing the spatial distributions of salt-solvent clusters in LHCE. The green stick model stands for DME molecule, light blue area for TFEO matrix, red ball-and-stick model for AGG+, blue ball-and-stick model for AGG, cyan ball-and-stick model for CIP, and dark grey ball-and-stick model for SSIP, while the black rectangular outline indicates the simulation boundary. **d**, Centre-of-mass (COM) radial distribution function (RDF) plots for the pairs of AGG *vs* TFEO, AGG+ *vs* TFEO, and DME *vs* TFEO. It is seen from **c** and **d** that AGG+ stay in the inner part of the network of salt-solvent clusters, while AGG and DME are mainly in the outer part. A fraction of DME molecules is completely dissolved into the TFEO matrix (i.e., free DME). **e**, Schematic for the spatial distributions of DME, AGG, and AGG+ in the LHCE. Green, blue, and red areas indicate Li$^+$-coordinated DME, AGG, and AGG+ regions, respectively. The green ovals represent free DME molecules that are miscible in the light blue TFEO matrix.

**Evolution of Micelle-like Structure in LHCE**

To provide insights into LHCE electrolyte design for further optimization, the factors to evolve the micelle-like structure are examined with a common LiFSI-1.2DME-2TFEO LHCE. A simple parameter that impacts salt-solvent solubility and solvent-diluent miscibility is temperature. The probability of different salt-solvent clusters as a function of temperature (0, 10, 25, 45, and 60 °C) were obtained through Raman deconvolution (Supplementary Fig. 6a-b) and MD (**Fig. 4a** and Supplementary Fig. 6c-d), both indicating a local AGG+ ratio peak at 25 °C in the 10-45 °C temperature range (Supplementary Fig. 6b and 6d and **Fig. 4b**). MD shows lower AGG+/AGG ratios and more fluctuations, likely caused by smaller cluster sizes and limited cluster numbers; smaller clusters have a larger surface (that contains more AGG) to volume ratio. Regardless, two temperature-dependent solubility/miscibility effects are revealed. First, as temperature increases, more DME migrate into the TFEO matrix (**Fig. 4c**). With fewer DME molecules coordinating with Li$^+$, anion-cation association increases inside the network of solvent-salt clusters. Second, DME dissolves more LiFSI as temperature increases, weakening the coordination strength between Li$^+$ and FSI$^-$ and causing decomposition of higher coordination aggregates, which is indirectly confirmed with binary HCEs (**Fig. 4d** and Supplementary Fig. 7). As a result, a "Goldilocks phenomenon" for the ratio of AGG+ is observed when these two effects are intertwined for the ternary LHCE. While the DME-TFEO miscibility effect is severe at extreme temperatures, it is mild in the 10-45 °C range, resulting in this Goldilocks phenomena of a local AGG+/AGG ratio peak at 25 °C (**Fig. 4b**). The competition of salt-solvent solubility and solvent-diluent miscibility in the micelle-like LHCE can also be reflected by the effect of the diluent concentration. The AGG+ ratio increases with diluent concentration, caused by more DME molecules mixing into the TFEO matrix, which is validated through Raman[9,10] (Supplementary Fig. 8) and MD (Supplementary Fig. 6d and Supplementary Fig. 9). Raman results reveal a breakpoint to this effect, as an excess of TFEO results in a reduction in higher aggregate clusters.

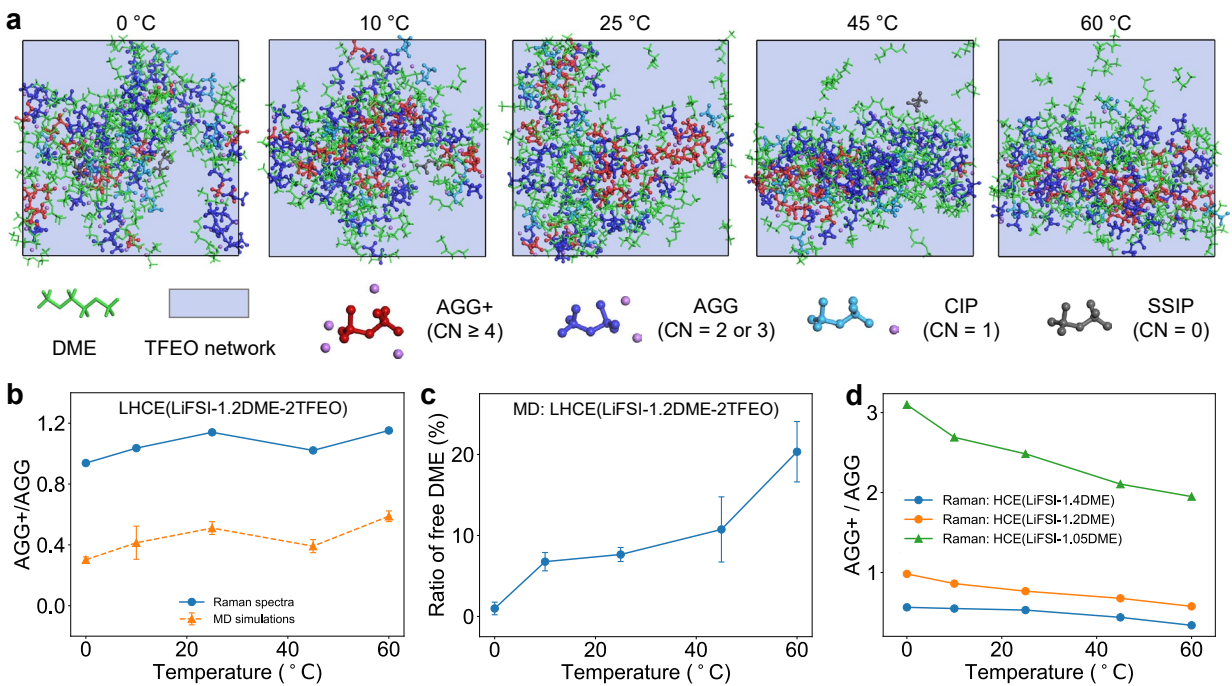

**Fig. 4 | Raman spectroscopy and MD simulations of LHCE and HCEs at various temperatures**. **a**, Snapshots taken from MD simulations showing the spatial distributions of the different cluster types (SSIP, CIP, AGG, and AGG+) as well as DME molecules, at different temperatures. **b**, AGG+/AGG ratios as functions of temperature calculated through Raman analysis (solid blue line) and MD statistics (dashed orange line) for LHCE. **c**, Ratio of free DME molecules (i.e., not coordinating with any $Li^+$) as a function of temperature for LHCE, calculated with MD. **d**, AGG+/AGG ratios as functions of temperature calculated through Raman analysis for HCEs (LiFSI-1.4DME, LiFSI-1.2DME, and LiFSI-1.05DME).

**Inspiring Formation Protocol for Practical Li-Metal Battery Application**

Conventional wisdom limits battery operation to 10-45 °C, since a high temperature leads to extensive side reactions and rapid capacity fade, while a low temperature limits lithium utilization due to slow kinetics.[34] However, operating at near extreme temperatures in a shortened time (e.g., formation cycles) permits forming a more stable initial SEI while minimizing detrimental temperature-driven impacts.

To observe how salt-solvent clusters affect initial SEI formation and cyclability, formation cycles were run at 10 °C, 25 °C, or 45 °C, followed by ageing cycles at 25 °C for LiFSI-1.2DME-2TFEO (**Fig. 5a-c**). The effects of temperature on overall cell capabilities are observed with the charge-discharge profiles during the first formation cycle, where lithium utilization (i.e., discharge capacity) and initial overpotentials follow Arrhenius temperature-dependence (**Fig. 5a**).[34] When equilibrated to 25 °C for ageing, discharge capacities and overpotentials are comparable (**Fig. 5b**), revealing that cell-level impacts are not substantial with the formation protocol at these different temperatures; rather, the primary impact of temperature is initial SEI formation driven by differences in LHCE microstructures. This is confirmed with cycle performance, where a 25 °C formation temperature outperformed 10 °C and 45 °C (**Fig. 5c**). This correlates to the increased AGG+ ratio at 25 °C (**Fig. 4b**) and thus an increase in salt-rendered SEI. Supplementary Fig. 10 shows that an increase in $Li^+$-$FSI^-$ coordination increases the reduction potential, easing anion decomposition at the anode surface.[8] Although both MD simulations and Raman peak deconvolution analyses confirm that a greater AGG+/AGG ratio is shown at 60 °C (**Fig. 4b**), the extent of macroscale impacts (e.g., extensive side reactions) would outweigh the benefit of improving electrolyte cluster statistics.

To confirm the impact of salt-solvent clusters on SEI formation, field emission scanning electron microscopy (FESEM) and X-ray photoelectron spectroscopy (XPS) with depth profiling were used to examine surface morphology and composition, respectively, of discharged anodes after formation cycles. The SEI is thin at 25 °C when compared to other temperatures (Supplementary Fig. 11). Peak positioning as a function of sputter time reveals monolithic behaviour for 25 °C with minimal transition in relative intensity (**Fig. 5d-f**). Additionally, there is minimal intensity in the organic carbon spectrum relative to inorganic components. For oxygen, the primary peak for 25 °C is $Li_2O$, whereas other temperatures are dominant in C-O and C=O, along with transitions in relative peak intensities over sputter time.[35] Initial SEIs formed at 10 °C shows more organic components than 45 °C due to an increase in probability of CIP structures (4.2% at 10 °C versus 2.5% at 45 °C, Supplementary Fig. 6b), driving poorer cycling performance. Hence, a relative reduction in organics versus inorganics (Supplementary Table 1) improves the initial SEI's chemical and mechanical stability. The formation of more inorganic SEI components would further suppress the decomposition of TFEO or DME (Supplementary Fig. 12). When examining FESEM results (**Fig. 5g-i**), stripping of Li at 10 °C and 45 °C resulted in non-uniform, porous surfaces, permitting active material consumption as cell operation continues. Comparatively, stripped Li foil at 25 °C is uniform and compact, primarily driven by the increased monolithic and inorganic-rich SEI. Therefore, the primary impact that the chosen temperatures had during formation cycles was in the variation of salt-solvent clusters in the micelle-like LHCE, which can be used to predict optimal formation cycle temperature.

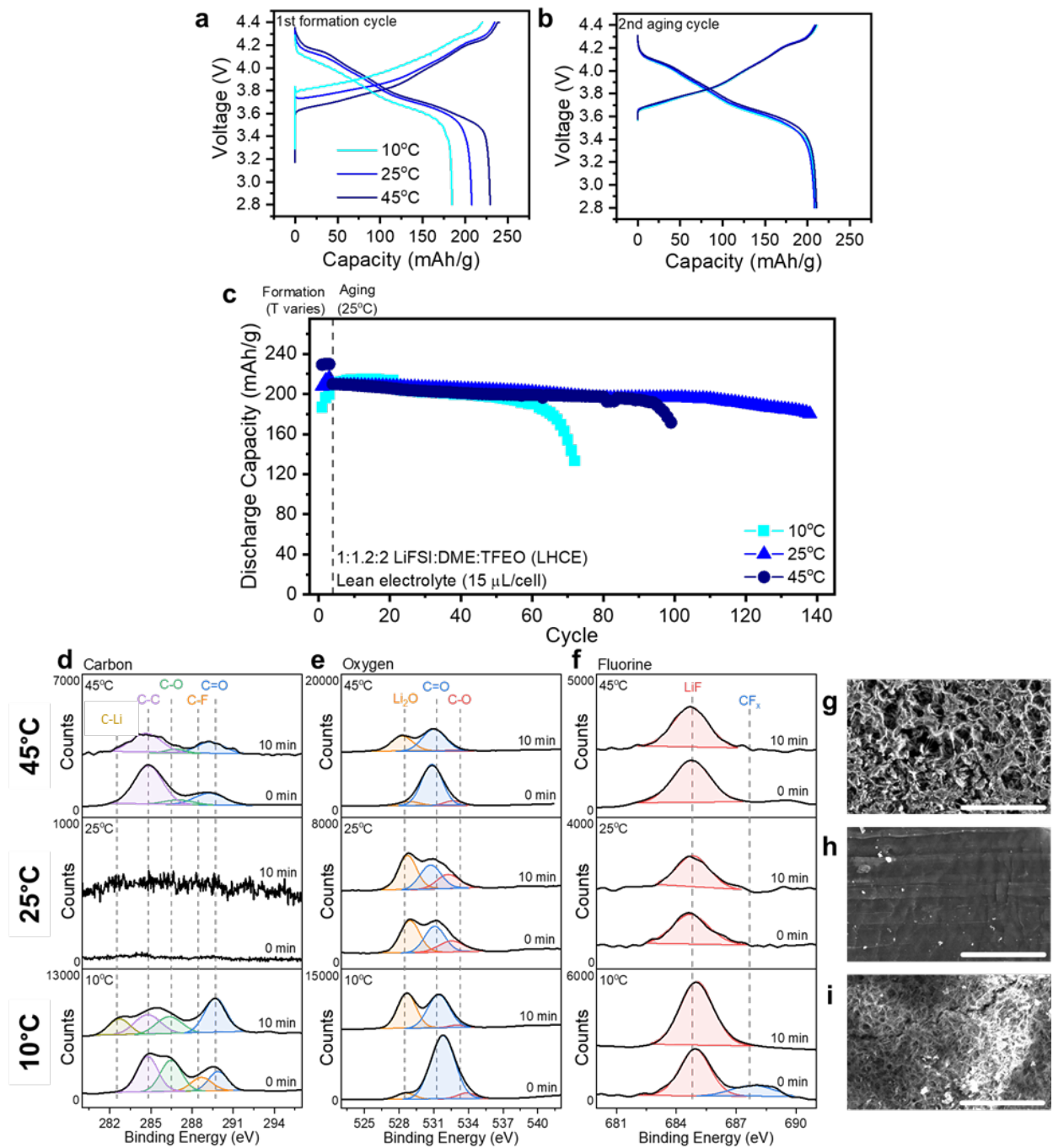

**Fig. 5 | Electrochemical performances of LHCE-based cells at various formation temperatures and corresponding SEI components and morphologies. a**, First formation cycle charge-discharge curves at several temperatures (10, 25, and 45 °C). **b**, Second ageing cycle charge-discharge curves at 25 °C following formation cycling at several temperatures (10, 25, and 45 °C). **c**, Cycling performance at 25 °C for different formation protocols. **d-f**, Surface and depth analysis of delithiated lithium foils with XPS of **d**, carbon peaks, **e**, oxygen peaks, and **f**, fluorine peaks after formation cycles at several temperatures (from bottom to top: 10, 25, and 45 °C). **g-i**,

FESEM images (scale bar of 50 μm) of delithiated lithium foil surface morphologies after formation cycles at **g**, 45 °C, **h**, 25 °C, and **i**, 10 °C.

**Control of Micelle-like Structures for LHCE Design**

Conventional micelles form beyond a critical concentration of surfactant, critical micelle concentration (CMC),[36] identified by conductivity trends.[37,38] **Fig. 6a** shows the ionic conductivity of LiFSI-1.2DME-$x$TFEO with increasing LiFSI concentration, where the slope changes at a critical "micelle-like" concentration ("CMC"), implying two different ionic conduction mechanisms. Below the "CMC", ionic conductivity increases with ion concentration in the uniformly dispersed solution. Above the "CMC", the ionic conductivity is mainly determined by the formation and connection of micelle-like structures. "CMC" is also identified in LiFSI-1.5DMC-$x$TTE (Supplementary Fig. 13).

Furthermore, the proposed micelle-like structures are identified by SAXS-WAXS (**Fig. 6b**). Electrolytes below the "CMC" ($x$ = 12 and 30) are comparable to the baseline (1.2DME-2TFEO), while electrolytes above "CMC" (x = 1 and 2) additionally peak at q ≈ 0.135 Å$^{-1}$ with a calculated diameter of ~ 47 Å ($2\pi/q$). The pair distance distribution function, P(r), shows an increasing peak in electrolytes above the "CMC" with an estimated radius of ~25 Å (**Fig. 6c**), suggesting formation of micelle-like structures consistent with the results of **Fig. 6a-b**. Similarly, an increase in TFEO results in Li$^+$-FSI$^-$ red-shifting in Raman spectra (Supplementary Fig. 8, $x$ = 12 and 30). As diluent concentration decreases, isolated micelles within the diluent matrix will connect into three-dimensional networks of salt-solvent clusters (**Fig. 2g-h**). This is validated by the structure factor, S(q), normalizing the SAXS data with respect to a dilute solution (Supplementary Fig. 14). Peak intensity increases with decreasing diluent concentration, suggesting more interactions among micelle-like structures. The presented results unify multiple recent discussions on varying ionic transport mechanisms,[39] ionic conductivity, and aggregation structures with LHCE compositions.[9,26] While knowing the "CMC" can guide the design of LHCEs, its structure and size depend on the chemistry and composition of the electrolyte, as well as external parameters (e.g., temperature).

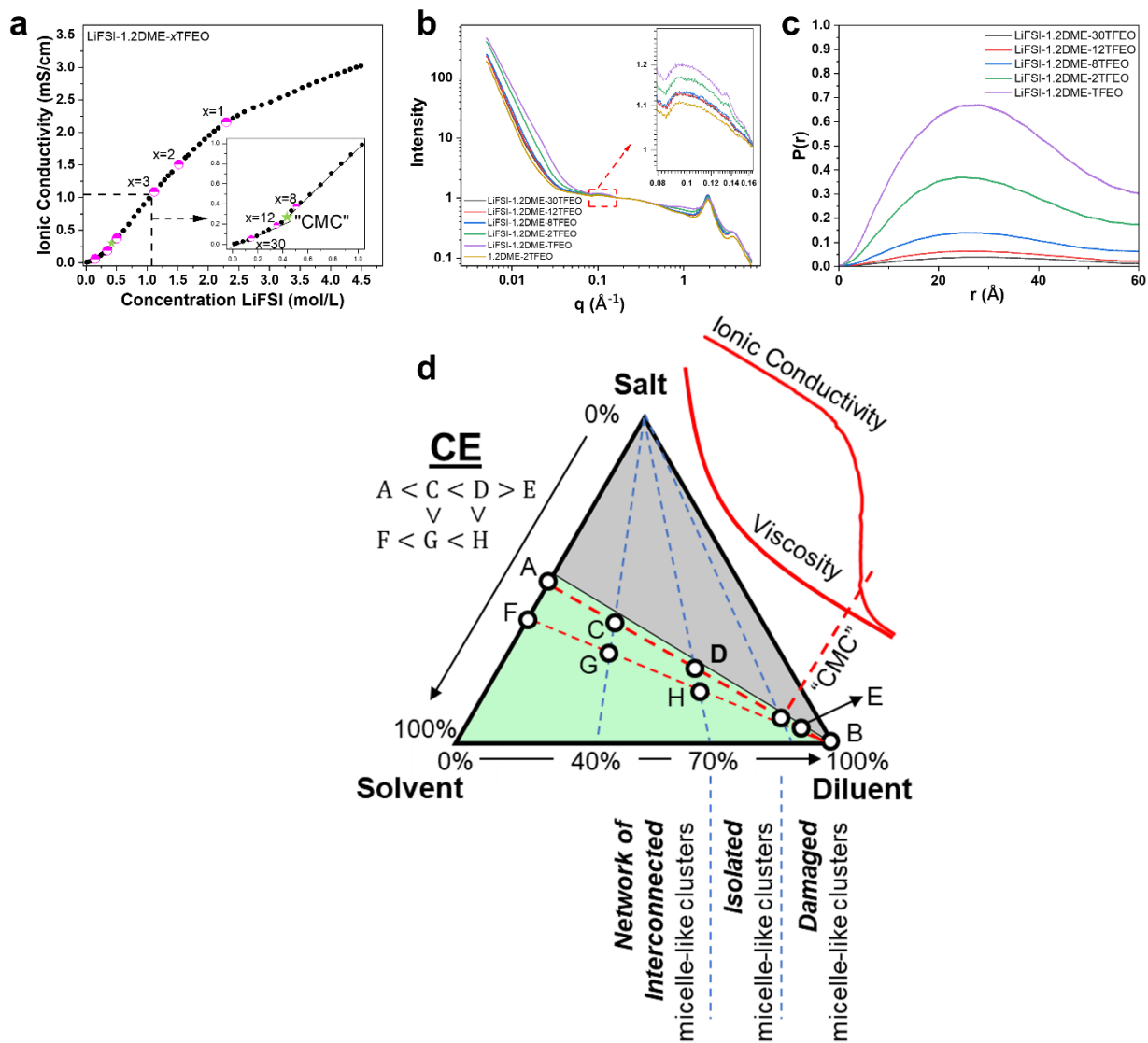

**Fig. 6 | Features of micelle-like structures in LHCE and rational LHCE design. a**, Ionic conductivities of LiFSI-1.2DME-$x$TFEO solutions as a function of LiFSI salt concentration, acquired between 21-22 °C. Specific solutions (pink circles), as well as the "CMC" (in the inset as a green star) are noted. **b**, SAXS-WAXS patterns acquired at 25 °C with a small q range of LiFSI-1.2DME-$x$TFEO solutions ($x$ = 1, 2, 8, 12 and 30) compared to 1.2DME-2TFEO co-solvent in the inset. **c**, Pair distance distribution function, P(r), of LiFSI-1.2DME-$x$TFEO solutions ($x$ = 1, 2, 8, 12 and 30) derived from SAXS patterns. **d**, Schematic showing LHCE design criteria according to ternary phase diagram. Coulombic efficiencies (CE) are ranked for each position shown in the phase diagram. Ionic conductivity and viscosity (red) are exemplified along line A-B. "CMC" is noted based on the ionic conductivity plot.

The ternary phase diagram (**Fig. 6d**) and the understanding of micelle-like structures illustrate the proposed design criteria for high-performance LHCE:

First, the concentration of diluent should be optimized by balancing macroscale (e.g., viscosity and ionic conductivity) and microscale properties (e.g., micelle-like structures). In general, the continuous addition of diluent reduces viscosity versus HCEs[9,40,41] and leads to an increase in local salt concentration. The gradual increase in local salt concentration with diluent is driven by the formation of the micelle-like structures and "CMC". By increasing the amount of diluent from point A to B (**Fig. 6d**), the local salt concentration first increases and then decreases (Supplementary Fig. 8), accompanied by transition points in the microstructure. Going from point C to point "CMC", the micelle-like microstructure forms and evolves from an interconnected network (C-D) to isolated clusters (D-"CMC") (**Fig. 2g-h**). This coordinates with an increase in local salt concentration and reduction in ionic conductivity (**Fig. 6a** and Supplementary Fig. 13). Continuing beyond "CMC" (e.g., point E), the micelle-like microstructure is damaged and reduces local salt concentration. Therefore, comprehensively considering the microstructures and macroscale properties, the diluent accounts for 40%-70% by mole (**Fig. 6d**), depending on the chemistries chosen (Supplementary Fig. 15a-b[9]).

Second, the electrolyte composition should be close to the "solubility line" to extend local salt concentration. This supports increased AGG+ formation, which forms salt-derived SEI, along with a higher CE value. In **Fig. 6d**, CE values at C and D should be higher than G and H, respectively, due to higher initial salt concentration in equivalent solvent-diluent solutions. Including the knowledge of optimizing diluent in criterium (a), CE values near D should be the highest due to highest local salt concentration and viable micelle-like microstructure. These design criteria are well validated with the LiFSI-DME-TFEO (Supplementary Fig. 15a) and the LiFSI-DMC-TTE systems (Supplementary Fig. 15b). For the LiFSI-DMC-TTE system, with the guidance shown in **Fig. 6d**, a CE above 99% was consistently achieved (Supplementary Fig. 15b and Supplementary Fig. 15). Saturated LiFSI-1.5DMC-3.07TTE is instable over longer cycling, though it accomplishes 99.59% CE for short-term Li∥Cu testing. LiFSI-2.2DMC-4.5TTE, with the second highest CE of 99.54%, improves upon reported LHCE systems (Supplementary Table 2, Supplementary Fig. 17).

Third, a salt/solvent system with higher salt solubility in solvents is preferred (Supplementary Fig. 15c).[42] This is exemplified by an increase in initial lithium hexafluorophosphate ($LiPF_6$) salt concentration in EC: ethyl methyl carbonate (EMC, 1:2 by mol) compared to LiPF$_6$ in DME. Additionally, the salt solubility should differ slightly when operating temperature varies, ensuring the electrolyte formulation always lies close to the "solubility line" in a wider temperature range. Understanding the necessary operation parameters and how that impacts the micelle-like structures is paramount for a viable large-scale battery in varied conditions.

In summary, a ternary phase diagram for the design of LHCE is proposed based on salt-solvent solubility and solvent-diluent miscibility. Salt-solvent clusters in LHCE exhibit micelle-like behaviour. A salt concentration gradient naturally forms in a micelle-like cluster, through which the ion-pair aggregates get more localized due to accumulation of solvent as a surfactant at salt network/diluent matrix interfaces. The micelle-like structure is also influenced by the temperature. In an exemplary LHCE of LiFSI-1.2DME-2TFEO, a localized peak ratio of AGG+ is seen at 25 °C, confirmed with both Raman and MD simulations, inspired a formation protocol that improved initial SEI composition and morphology, and extended the cyclability. In the LiFSI-DMC-TTE system, an unprecedented CE above 99.5% is accomplished, optimized by compensating microstructures (e.g., micelle-like structures and network versus isolated clusters) with macroscale properties (e.g., ionic conductivity). This work proposes methods of controlling the micelle-like structure in LHCE, supported by SAXS and Raman characterization, MD simulations as well as electrochemical measurements, for higher performing practical batteries. From here, the impacts of electrolyte component choices in LHCEs to control salt-solvent cluster size, shape, and composition, as well as external parameters chosen during operation (e.g., temperature) can be optimized to extend anode stability and cyclability of high-energy batteries.

## Methods

**Materials synthesis.** Lithium foil (50 μm, China Energy Lithium Co.) was punched into ~1.43 cm diameter disks and rolled onto stainless steel spacers. NMC811 cathode coated on aluminium foil (Pacific Northwest National Lab, PNNL) was stored in a glovebox and punched into 1.27 cm diameter disks. NMC811 has a practical capacity of 4.34 mAh/cm$^2$. Celgard® 2325, cut at ~1.59 cm diameter, was used as a separator. Different ether-based electrolytes were formulated and labelled as such: low-concentration electrolyte (LCE) as 1:9 molar ratio lithium bis(fluorosulfonyl)imide (LiFSI, Nippon Shokubai Co.) in 1,2-dimethoxyethane (DME, Sigma-Aldrich); high-concentration electrolyte (HCE) as various molar ratios of LiFSI in DME (1:1.4, 1:1.2, 1:1.05); localized high-concentration electrolyte (LHCE) incorporating diluent tris(2,2,2-

trifluoroethyl)orthoformate (TFEO, SynQuest Laboratories) in 1:1.2:x LiFSI:DME:TFEO molar ratios (x = 1, 2, 8, 12, and 30). Prior to coin cell assembly, CR2032 components, spacers, and spring (MTI Corporation) were ultrasonicated in ethanol for 15 minutes, followed by DI water for 15 minutes. Materials were then dried at 60 ºC under vacuum below -75 kPa for at least 8 hours and held under vacuum in the argon (Ar)-filled glovebox antechamber for at least 8 hours prior to loading into glovebox. Aluminium cladded foil and positive-side case were used on the cathode-side to reduce LiFSI corrosion to the stainless-steel positive case.

**Electrochemical measurements.** All electrochemical experiments were done with CR2032-type cells (MTI). Galvanostatic cycling of Li‖NMC811 or Cu‖NMC811 full cells was done between 2.8 and 4.4 V operating window with C/10 charge and discharge rates for 3 formation cycles in environmental chambers set at 10 °C, 25 °C, or 45 °C, where 1C = 4.34 mA/cm$^2$. 15 μL of electrolyte was loaded into each cell. Cells were rested for 1 hour between each charge and discharge half-cycle. Upon completion of formation cycles, cells were placed into a 25 °C chamber and left at rest for 12 hours to allow thermal equilibration. Ageing cycles at C/10 charge and C/5 discharge rates were then run at 25 °C regardless of formation cycle temperature, with 15 minutes of rest between charge and discharge half-cycles. Ionic conductivity of electrolyte was measured in a glovebox at room temperature with a Model CM-30R Conductivity Meter (DKK-TOA).

Coulombic efficiency (CE) tests are referred to elsewhere using the modified Aurbach's method, "Method 3",[11] protocol with a formation capacity ($Q_T$) of 5 mAh/cm$^2$, cycling capacity ($Q_C$) of 1 mAh/cm$^2$, a cycling rate of 0.5 mA/cm$^2$, and number of cycles (n) of 10.[11] Li‖Cu cells were assembled inside an argon-filled glovebox (MBraun, H$_2$O < 1 ppm, O$_2$ < 1 ppm). 75 μL of electrolyte was added to each cell.

**Characterization**. Raman analysis was conducted by placing the solution (or solid salt) on a concave microscope slide (W. W. Grainger, Inc.), then sealing the slide with optical adhesive and a fused silica disk (Edmund Optics). The samples were exposed to various temperatures of 0 °C, 10 °C, 25 °C, 45 °C, and 60 °C for Raman measurements. Raman spectroscopy was accomplished using a HORIBA LabRAM HR Evolution (HORIBA Scientific) equipped with a 50 mW monochromatic 532 nm doubled Nd:YAG laser with -0.3 cm$^{-1}$ spectral resolution. Spatial resolution, with a 20x lens magnification, was between 0.5 and 1 μm. Spectra were processed with LabSpec V6.3.x (HORIBA Scientific). Spectra underwent a baseline correction and deNoise to remove broadening and background noise, respectively. Gauss peak fitting with a zeroed y-offset was used to deconvolute peaks of interest from spectra with $\geq 0.98$ $R^2$ regression. The area of each fitted peak was used to compare different peak contributions.

Post-mortem analysis of lithium foils was done after formation cycles at 10 °C, 25 °C, and 45 °C. Cells were decrimped after cycling, followed by electrode rinsing with DME and dried in vacuum prior to analysis. X-ray photoelectron spectroscopy (XPS) with a 1253.6 eV Mg (Kα) X-ray source was used to provide surface analysis with a PHI-5600 (Physical Electronics), along with an $Ar^+$ ion gun (2 kV, 1.2 μA) for sputter depth profiling. Sample charging was neutralized with a low-energy electron-gun. A vacuum transfer vessel (PHI Model 04-110) was used to prevent air exposure. Peak position calibration is referenced to adventitious C1s, C-C peak at 284.8 eV, or to LiF at 684.7 eV if there was limited carbon present. PHI MultiPak software (Physical Electronics) was used with a mixed Gauss-Lorentzian peak fitting with >80% Gauss for each XPS peak. A FEI Teneo field emission scanning electron microscope (FESEM) was used to observe surface morphology. Brief air exposure (< 1 minute) occurred when transferring electrodes into the FESEM chamber.

The small-angle X-ray scattering (SAXS) and wide-angle X-ray scattering (WAXS) measurements were conducted at the Soft Matter Interfaces beamline (12-ID) of the National Synchrotron Light Source II (NSLS-II) at Brookhaven National Laboratory. The liquid samples were loaded into Kapton capillaries with a diameter of 1.5 mm, which were then well sealed and mounted on the SMI sample stage. The scattered data were collected using a beam energy of 16.1 keV and beam size of 200 × 30 um. A Pilatus 1M area detector (Dectris, Switzerland) was used for SAXS. The detector, consisting of 0.172 mm square pixels in a 981 × 1043 array, was placed five meters downstream from the sample position. The WAXS data were collected with a PILATUS3 900 kW detector (Dectris, Switzerland), consisting of 0.172 mm square pixels in a 1475 × 619 array. To obtain a wide range of wave vector transfer (q), a series of 2D diffraction patterns were collected by rotating the WAXS detector on an arc with the sample-to-detector distance being 275 mm. Scattering patterns from each detector angle were stitched together using home-developed software. Then, both SAXS and WAXS 2D scattering patterns were reduced to 1D scattering intensity, I(q), by circular average. The q is wave vector transfer, $q = (4\pi/\lambda) \sin(\theta)$, where λ = 0.77 Å and 2θ is the wavelength of the incident x-ray beam and the scattering angle, respectively.

**MD simulations.** All MD simulations were conducted using the Forcite module in the Materials Studio (MS) 2020.[43] The COMPASS III force field was used along with optimized atom types and charges, which were all taken from previous works,[44] except for that the charges of $Li^+$ and $FSI^-$ from the salts are scaled by 0.7 to properly account for the ion-ion and ion-dipole interactions. The representative atom types and charges are shown in Supplementary Fig. 18. In terms of the cation-anion/solvent/diluent interactions and the density, ion conductivity, $Li^+$ coordination, ion paring and aggregation ratios in both high and low concentration electrolytes, the scaling factor of 0.7 either gave similar or better results (closer agreement with experiments[3,45] and/or simulation results based on polarizable force field,[3,46] density functional theory[47]), when compared with the scaling factor of 0.8, as discussed in our previous publication[48] and Supplementary Information (Supplementary Figs. 19-20 and Supplementary Tables 3-6). Especially switching the scaling factor from 0.8 to 0.7 would result in a decrease in $FSI^-$ coordination with $Li^+$ (through its O atoms) by almost 1 in 1M LiFSI-9DME electrolyte. The difference is much less for $LiPF_6$ pairs in EC-EMC solvents. As this is critical for solvation structures, it requires careful tests or using the systematic Molecular Dynamics Electronic Continuum (MDEC) model,[49,50] which gave an optimal scaling factor (0.73) for LiFSI in DME.

It is challenging to obtain equilibrium heterogenous liquid structures in LHCE. We approached this by considering different initial structures: (a) immersing a LiFSI salt cluster in mixed solvent/diluent; (b) comparing salt/solvent clusters at different sizes in TFEO diluent; and (c) randomly mixing/packing all species (LiFSI, DME and TFEO) through the Amorphous module in MS 2020.[43] After 20 ns dynamics, the initial structures with salt-solvent clusters (b) showed lower energy than (a) and (c), and the initial structure with the lowest average energy was used to mimic LHCE structure (Supplementary Fig. 20).

The electrolyte systems (Supplementary Table 7) were subjected to three stages of constant number, pressure, and temperature (NPT) simulations, including a 2.0 ns pre-equilibrium run at room temperature, a long equilibrium run at desired temperatures (0 °C, 10 °C, 25 °C, 45 °C, and 60 °C), and a 4 ns production run to obtain statistics. The LHCEs (LIFSI-1.2DME-2TFEO and LiFSI-1.2DME-8TFEO) requires a longer equilibrium run of 16.0 ns, compared to 4 ns equilibrium run that is needed for HCEs (LiFSI-1.4DME and LiFSI-1.2DME) and LCE (LiFSI-9DME). The Nose-Hoover method[51] and Berendsen method[52] were used to control the temperature and pressure, respectively. For the LiFSI crystal simulations, equilibrium runs were performed for 1.0 ns followed by production runs for 1.0 ns at 25 °C. For both the LiFSI-TFEO and DME-TFEO mixture systems, the NPT simulations were conducted for 22.0 ns at 25 °C.

**MD-based coordination number (CN) analyses.** The statistics of the CN and the subsequent categorized aggregate ratios were analysed through our home-made perl and Python scripts, which are available upon request (Supplementary Fig. 21-25). Through the time evolution $CN(t)$, the time averaged $\langle CN(t) \rangle$ from the beginning of the production run to time $t$ was calculated. All the reported values are averaged for 4 ns production run, $\langle CN(4\ ns) \rangle$. We defined the error bar as the difference between the maximum and minimum in the time-averaged values, $error = \pm(\ max\{\langle CN(t)\rangle\} - min\{\langle CN(t)\rangle\})$, from 2 ns to 4 ns during production runs (see Supplementary Figs. 23-26 for time evolution and running averages of CN and ratios of aggregates). The error bars are generally small in homogenous LCE and HCE and become larger in heterogenous LCHE. Thus, we run additional 10 ns NPT dynamics for LHCE for further validation (Supplementary Fig. 27). The conclusion holds considering the error bar, including the occurrence of a peak value of AGG+/AGG at room temperature, which show larger scattering. This is likely due to the smaller cluster sizes in MD simulations compared to experiments. In our coordination number (CN) analyses, FSI⁻ (or DME, TFEO) and Li⁺ are considered being coordinated with each other if the Li⁺ ion falls within 2.8 Å from any of the O, N and F atoms in the FSI⁻ anion (or DME, TFEO). It is seen in the radial distribution function (RDF) plots (Supplementary Fig. 1) that the value of 2.8 Å is close to the first minimum after the primary peak (~3.0 Å), which is often considered as the first coordination shell in literature.[53] Our analyses showed that the same trends can be obtained in terms of the ratios of the salt-solvent clusters (SSIP, CIP, AGG, and AGG+) when using other cutoff values (2.4 Å or 3.2 Å), as shown in Supplementary Fig. 28. DME molecule in LHCE is considered as free DME when it is not coordinating with Li⁺ ions (i.e., none of its ether oxygen atoms are within 2.8 Å of any Li⁺).

**Density functional theory (DFT) calculations.** All DFT calculations were conducted using the Gaussian 09 code.[54] The double hybrid functional M06-2X[55] and the basis set 6-31+G** along with the D3 dispersion correction[56] were used. The implicit SMD model[57] and the dielectric constant of 7.2 were used to account for the solvation environment when calculating the reduction and oxidation potentials.

## Acknowledgements

The authors at Idaho National Laboratory (INL), Brookhaven National Laboratory (BNL), and Pacific Northwest National Laboratory (PNNL) thank the financial support by the Assistant Secretary for Energy Efficiency and Renewable Energy, Office of Vehicle Technologies of the U.S. Department of Energy through the Advanced Battery Materials Research Program (Battery500 Consortium). Q. Wu and Y. Qi thank NASA for financial support (grant no. 80NSSC21M0107). INL is operated by Battelle Energy Alliance under Contract Nos. DE-AC07-05ID14517 for the U.S. Department of Energy. PNNL is operated by Battelle under Contract No. DE-AC05-76RLO1830 for the U.S. Department of Energy. Additional financial support was provided by the Micron School of Materials Science and Engineering at Boise State University. The authors acknowledge the Atomic Films Laboratory at Boise State University for the use of the PHI-5600 XPS system. This research also used resources of the Center for Functional Nanomaterials and the SMI beamline (12-ID) of the National Synchrotron Light Source II, both supported by U.S. DOE Office of Science Facilities at BNL under Contract No. DE-SC0012704. The authors thank Elton Graugnard, JD Hues, and Jake Soares for support with XPS, Nicholas Bulloss for support with FESEM, and Paul H. Davis for support with Raman as well as Sha Tan from BNL for electrolyte sample preparation.

## Author contributions

B.L. and Y.Q. conceived the original idea and designed the experiments. Q.W. and Y.Q. conducted all MD simulations and DFT calculations, as well as computational analyses. C.M.E and B. L. collected and processed Raman and FESEM data. C.M.E. and N.G. prepared and cycled coin cells. X.C. prepared electrolytes and cycled Coulombic efficiency cells. H.Z and C.M.E. collected and processed XPS results. Y. Z., B. L., Y. Q., E. H., X. Q. Y. and J. L. collected and processed SAXS-WAXS results. C.M.E., Q.W., Y.Q., and B.L. wrote the manuscript. All authors contributed to the discussions and revisions of the manuscript.

## Conflict of Interest

The authors declare no conflict of interest.

## Additional Information

**Supplementary information** is available for this paper.

## References

1       Wang, Z. *et al.* Structural regulation chemistry of lithium ion solvation for lithium batteries. *EcoMat*, 1-24, doi:10.1002/eom2.12200 (2022).

2       Cheng, H. *et al.* Emerging Era of Electrolyte Solvation Structure and Interfacial Model in Batteries. *ACS Energy Letters* **7**, 490-513, doi:10.1021/acsenergylett.1c02425 (2022).

3       Qian, J. *et al.* High rate and stable cycling of lithium metal anode. *Nature Communications* **6**, doi:10.1038/ncomms7362 (2015).

4       Wang, J. *et al.* Superconcentrated electrolytes for a high-voltage lithium-ion battery. *Nature Communications* **7**, 1-9, doi:10.1038/ncomms12032 (2016).

5       Yamada, Y. & Yamada, A. Review—Superconcentrated Electrolytes for Lithium Batteries. *Journal of The Electrochemical Society* **162**, A2406-A2423, doi:10.1149/2.0041514jes (2015).

6       Suo, L., Hu, Y. S., Li, H., Armand, M. & Chen, L. A new class of Solvent-in-Salt electrolyte for high-energy rechargeable metallic lithium batteries. *Nature Communications* **4**, 1-9, doi:10.1038/ncomms2513 (2013).

7       Cao, X., Jia, H., Xu, W. & Zhang, J.-G. Review—Localized High-Concentration Electrolytes for Lithium Batteries. *Journal of The Electrochemical Society* **168**, 010522-010522, doi:10.1149/1945-7111/abd60e (2021).

8       Cao, X. *et al.* Monolithic solid-electrolyte interphases formed in fluorinated orthoformate-based electrolytes minimize Li depletion and pulverization. *Nat Energy* **4**, 796-805, doi:10.1038/s41560-019-0464-5 (2019).

9       Cao, X. *et al.* Optimization of fluorinated orthoformate based electrolytes for practical high-voltage lithium metal batteries. *Energy Storage Mater* **34**, 76-84, doi:10.1016/j.ensm.2020.08.035 (2021).

10      Li, T. *et al.* Stable Anion-Derived Solid Electrolyte Interphase in Lithium Metal Batteries. *Angew Chem Int Edit*, doi:10.1002/anie.202107732 (2021).

11  Adams, B. D., Zheng, J. M., Ren, X. D., Xu, W. & Zhang, J. G. Accurate Determination of Coulombic Efficiency for Lithium Metal Anodes and Lithium Metal Batteries. *Adv Energy Mater* **8**, doi:10.1002/aenm.201702097 (2018).

12  Jia, H. P. *et al.* Controlling Ion Coordination Structure and Diffusion Kinetics for Optimized Electrode-Electrolyte Interphases and High-Performance Si Anodes. *Chem Mater* **32**, 8956-8964, doi:10.1021/acs.chemmater.0c02954 (2020).

13  Peng, X. D., Lin, Y. K., Wang, Y., Li, Y. J. & Zhao, T. S. A lightweight localized high-concentration ether electrolyte for high-voltage Li-Ion and Li-metal batteries. *Nano Energy* **96**, doi:10.1016/j.nanoen.2022.107102 (2022).

14  Wang, Y. D. *et al.* Enhanced Sodium Metal/Electrolyte Interface by a Localized High-Concentration Electrolyte for Sodium Metal Batteries: First-Principles Calculations and Experimental Studies. *Acs Appl Energ Mater* **4**, 7376-7384, doi:10.1021/acsaem.1c01573 (2021).

15  Wang, N. *et al.* Stabilized Rechargeable Aqueous Zinc Batteries Using Ethylene Glycol as Water Blocker. *Chemsuschem* **13**, 5556-5564, doi:10.1002/cssc.202001750 (2020).

16  Xue, R. F. *et al.* Highly reversible zinc metal anodes enabled by a three-dimensional silver host for aqueous batteries. *J Mater Chem A* **10**, 10043-10050, doi:10.1039/d2ta00326k (2022).

17  Du, X. Q. & Zhang, B. A. Robust Solid Electrolyte Interphases in Localized High Concentration Electrolytes Boosting Black Phosphorus Anode for Potassium-Ion Batteries. *Acs Nano* **15**, 16851-16860, doi:10.1021/acsnano.1c07414 (2021).

18  Qin, L. *et al.* Pursuing graphite-based K-ion $O_2$ batteries: a lesson from Li-ion batteries. *Energ Environ Sci* **13**, 3656-3662, doi:10.1039/d0ee01361g (2020).

19  Piao, N. *et al.* Countersolvent Electrolytes for Lithium-Metal Batteries. *Advanced Energy Materials* **10**, 1-9, doi:10.1002/aenm.201903568 (2020).

20  Qian, K., Winans, R. E. & Li, T. Insights into the Nanostructure, Solvation, and Dynamics of Liquid Electrolytes through Small-Angle X-Ray Scattering. *Advanced Energy Materials* **11**, 1-22, doi:10.1002/aenm.202002821 (2021).

21  Su, C. C. *et al.* Solvating power series of electrolyte solvents for lithium batteries. *Energy and Environmental Science* **12**, 1249-1254, doi:10.1039/c9ee00141g (2019).

22  Cao, X., Zhang, J.-G. & Xu, W. Electrolyte for Stable Cycling of Rechareable Aklali Metal and Alkali Ion Batteries. United States patent (2019).

23  McBain, J. W. Mobility of Highly-Charged Micelles. *Transactions of the Faraday Society* **9**, 99 (1913).

24  McClements, D. J. Nanoemulsions versus microemulsions: terminology, differences, and similarities. *Soft Matter* **8**, 1719-1729, doi:10.1039/c2sm06903b (2012).

25  Zhao, Y. *et al.* A Micelle Electrolyte Enabled by Fluorinated Ether Additives for Polysulfide Suppression and Li Metal Stabilization in Li-S Battery. *Frontiers in Chemistry* **8**, 1-9, doi:10.3389/fchem.2020.00484 (2020).

26  Ren, F. *et al.* Solvent–Diluent Interaction-Mediated Solvation Structure of Localized High-Concentration Electrolytes. *ACS Applied Materials and Interfaces* **14**, 4211-4219, doi:10.1021/acsami.1c21638 (2022).

27  Beltran, S. P., Cao, X., Zhang, J. G. & Balbuena, P. B. Localized High Concentration Electrolytes for High Voltage Lithium-Metal Batteries: Correlation between the Electrolyte Composition and Its Reductive/Oxidative Stability. *Chem Mater* **32**, 5973-5984, doi:10.1021/acs.chemmater.0c00987 (2020).

28  Genovese, M. *et al.* Hot Formation for Improved Low Temperature Cycling of Anode-Free Lithium Metal Batteries. *J Electrochem Soc* **166**, A3342-A3347, doi:10.1149/2.0661914jes (2019).

29  Yoshida, H. & Matsuura, H. Density functional study of the conformations and vibrations of 1,2-dimethoxyethane. *Journal of Physical Chemistry A* **102**, 2691-2699, doi:10.1021/jp9800766 (1998).

30  Cote, J. F. *et al.* Dielectric constants of acetonitrile, gamma-butyrolactone, propylene carbonate, and 1,2-dimethoxyethane as a function of pressure and temperature. *Journal of Solution Chemistry* **25**, 1163-1173, doi:Doi 10.1007/Bf00972644 (1996).

31  Pham, T. A., Kweon, K. E., Samanta, A., Lordi, V. & E., P. J. Solvation and Dynamics of Sodium and Potassium in Ethylene Carbonate from ab Initio Molecular Dynamics Simulations. *The Journal of Physical Chemistry C* **121**, 21913-21920, doi:10.1021/acs.jpcc.7b06457 (2017).

32  Kerner, M., Plylahan, N., Scheers, J. & Johansson, P. Thermal stability and decomposition of lithium bis(fluorosulfonyl)imide (LiFSI) salts. *RSC Advances* **6**, 23327-23334, doi:10.1039/c5ra25048j (2016).

33  Suo, L., Zheng, F., Hu, Y. S. & Chen, L. FT-Raman spectroscopy study of solvent-in-salt electrolytes. *Chinese Physics B* **25**, doi:10.1088/1674-1056/25/1/016101 (2015).

34	Wang, J. *et al.* Improving cyclability of Li metal batteries at elevated temperatures and its origin revealed by cryo-electron microscopy. *Nature Energy* **4**, 664-670, doi:10.1038/s41560-019-0413-3 (2019).

35	Sun, B. *et al.* At the polymer electrolyte interfaces: the role of the polymer host in interphase layer formation in Li-batteries. *Journal of Materials Chemistry A* **3**, 13994-14000, doi:10.1039/C5TA02485D (2015).

36	Nagarajan, R. in *Structure-Performance Relationships in Surfactants* (eds Esumi K & Ueno M) 1-89 (Marcel Dekker, 2003).

37	Pal, A. & Chaudhary, S. Ionic liquids effect on critical micelle concentration of SDS: Conductivity, fluorescence and NMR studies. *Fluid Phase Equilibr* **372**, 100-104, doi:10.1016/j.fluid.2014.03.024 (2014).

38	Perez-Rodriguez, M. *et al.* A comparative study of the determination of the critical micelle concentration by conductivity and dielectric constant measurements. *Langmuir* **14**, 4422-4426, doi:10.1021/la980296a (1998).

39	Yamada, Y., Wang, J., Ko, S., Watanabe, E. & Yamada, A. Advances and issues in developing salt-concentrated battery electrolytes. *Nature Energy* **4**, 269-280, doi:10.1038/s41560-019-0336-z (2019).

40	Chen, S. R. *et al.* High-Efficiency Lithium Metal Batteries with Fire-Retardant Electrolytes. *Joule* **2**, 1548-1558, doi:10.1016/j.joule.2018.05.002 (2018).

41	Ren, X. D. *et al.* Localized High-Concentration Sulfone Electrolytes for High-Efficiency Lithium-Metal Batteries. *Chem-Us* **4**, 1877-1892, doi:10.1016/j.chempr.2018.05.002 (2018).

42	Su, L. S. *et al.* Uncovering the Solvation Structure of LiPF6-Based Localized Saturated Electrolytes and Their Effect on LiNiO2-Based Lithium-Metal Batteries. *Adv Energy Mater*, doi:10.1002/aenm.202201911 (2022).

43	Knap, V. *et al.* Reference Performance Test Methodology for Degradation Assessment of Lithium-Sulfur Batteries. *J Electrochem Soc* **165**, A1601-A1609, doi:10.1149/2.0121809jes (2018).

44	Akkermans, R. L. C., Spenley, N. A. & Robertson, S. H. compass iii: automated fitting workflows and extension to ionic liquids. *Mol Simulat* **47**, 540-551, doi:10.1080/08927022.2020.1808215 (2021).

45  von Cresce, A. & Xu, K. Preferential Solvation of Li+ Directs Formation of Interphase on Graphitic Anode. *Electrochem Solid St* **14**, A154-A156, doi:10.1149/1.3615828 (2011).

46  Borodin, O. & Smith, G. D. Quantum Chemistry and Molecular Dynamics Simulation Study of Dimethyl Carbonate: Ethylene Carbonate Electrolytes Doped with LiPF6. *J Phys Chem B* **113**, 1763-1776, doi:10.1021/jp809614h (2009).

47  Borodin, O. *et al.* Competitive lithium solvation of linear and cyclic carbonates from quantum chemistry. *Phys Chem Chem Phys* **18**, 164-175, doi:10.1039/c5cp05121e (2016).

48  Wu, Q. S., McDowell, M. T. & Qi, Y. Effect of the Electric Double Layer (EDL) in Multicomponent Electrolyte Reduction and Solid Electrolyte Interphase (SEI) Formation in Lithium Batteries. *J Am Chem Soc*, 2473-2484, doi:10.1021/jacs.2c118072473J (2023).

49  Liu, H. *et al.* Ultrahigh coulombic efficiency electrolyte enables Li||SPAN batteries with superior cycling performance. *Mater Today* **42**, 17-28, doi:10.1016/j.mattod.2020.09.035 (2021).

50  Park, C. *et al.* Molecular simulations of electrolyte structure and dynamics in lithium-sulfur battery solvents. *J Power Sources* **373**, 70-78, doi:DOI 10.1016/j.jpowsour.2017.10.081 (2018).

51  Nose, S. A Unified Formulation of the Constant Temperature Molecular-Dynamics Methods. *J Chem Phys* **81**, 511-519, doi:10.1063/1.447334 (1984).

52  Berendsen, H. J. C., Postma, J. P. M., van Gunsteren, W. F., DiNola, A. & Haak, J. R. Molecular dynamics with coupling to an external bath. *The Journal of Chemical Physics* **81**, 3684-3690, doi:10.1063/1.448118 (1984).

53  Toxvaerd, S. & Dyre, J. C. Role of the first coordination shell in determining the equilibrium structure and dynamics of simple liquids. *The Journal of Chemical Physics* **135**, 134501-134501, doi:10.1063/1.3643123 (2011).

54  Gaussian 09, Revision D.01 (Gaussian, Inc., Wallingford CT, 2016).

55  Zhao, Y. & Truhlar, D. G. The M06 suite of density functionals for main group thermochemistry, thermochemical kinetics, noncovalent interactions, excited states, and transition elements: two new functionals and systematic testing of four M06-class functionals and 12 other functionals. *Theor Chem Acc* **120**, 215-241, doi:10.1007/s00214-007-0310-x (2008).

56	Grimme, S., Ehrlich, S. & Goerigk, L. Effect of the Damping Function in Dispersion Corrected Density Functional Theory. *J Comput Chem* **32**, 1456-1465, doi:10.1002/jcc.21759 (2011).

57	Marenich, A. V., Cramer, C. J. & Truhlar, D. G. Universal Solvation Model Based on the Generalized Born Approximation with Asymmetric Descreening. *J Chem Theory Comput* **5**, 2447-2464, doi:10.1021/ct900312z (2009).

Supplementary information for

# Localized High-Concentration Electrolytes Get More Localized Through Micelle-Like Structures

Corey M. Efaw,[1,2,#] Qisheng Wu,[3,#] Ningshengjie Gao,[1] Yugang Zhang,[4] Haoyu Zhou,[2] Kevin Gering,[1] Michael F. Hurley,[2] Hui Xiong,[2] Enyuan Hu,[5] Xia Cao,[6] Wu Xu,[6] Ji-Guang Zhang,[6] Eric J. Dufek,[1] Jie Xiao,[6,7] Xiao-Qing Yang,[5] Jun Liu,[6,7] Yue Qi,[3,*] and Bin Li[1,2,*]

[1]Energy and Environmental Science and Technology, Idaho National Laboratory, Idaho Falls, ID 83415, USA

[2]Micron School of Materials Science and Engineering, Boise State University, Boise, ID 83725, USA

[3]School of Engineering, Brown University, Providence, RI 02912, USA

[4]Center for Functional Nanomaterials, Brookhaven National Laboratory, Upton, New York 11973, USA

[5]Chemistry Division, Brookhaven National Laboratory, Upton, New York 11973, USA

[6]Materials Science and Engineering Department, University of Washington, Seattle, WA, 98105 USA

[7]Energy and Environment Directorate, Pacific Northwest National Laboratory, Richland, WA, 99252 USA

#These authors contributed equally to this work

*Corresponding Authors: bin.li@inl.gov and yueqi@brown.edu

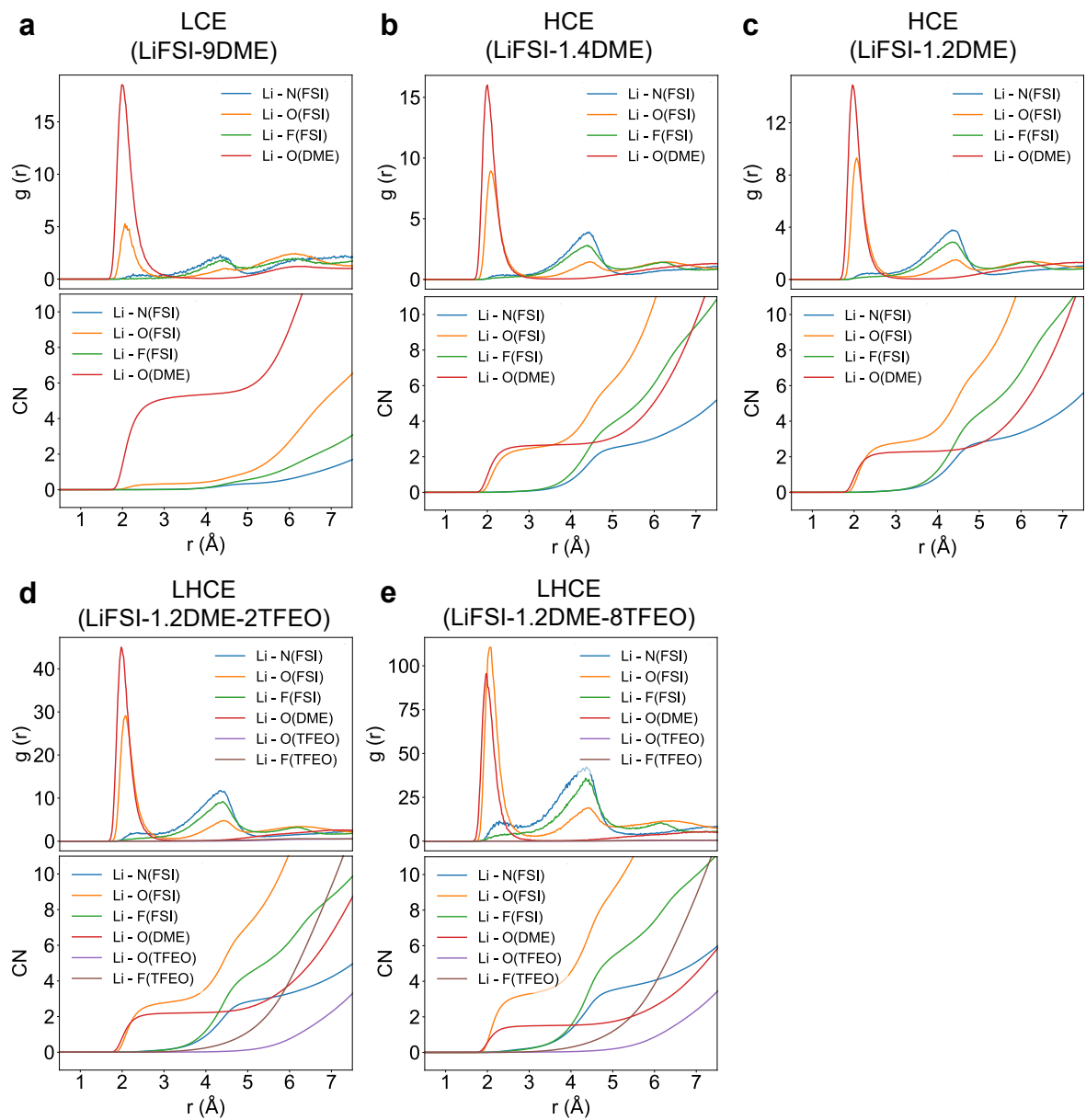

**Supplementary Fig. 1 | Radial distribution function (RDF) and coordination number (CN) plots. a-c**, Time averaged RDF and CN as functions of distances between Li$^+$ and N, O and F atoms in FSI$^-$ and DME molecules for **a**, LCE (LiFSI-9DME), **b**, HCE (LiFSI-1.4DME), and **c**, HCE (LiFSI-1.2DME) obtained through molecular dynamics (MD) simulations and analysis at 25 °C. In LCE, the Li$^+$ solvation shell is mainly composed of O atoms from DME, while both FSI$^-$ and DME contribute significantly to the Li$^+$ solvation shell in HCEs. **d-e**, RDF and CN as functions of distances between Li$^+$ and N, O and F atoms in FSI$^-$, DME and TFEO molecules for, **d**, LHCE (LiFSI-1.2DME-2TFEO) and, **e**, LHCE (LiFSI-1.2DME-8TFEO) obtained through MD simulations and analysis at 25 °C. In LHCEs, the Li$^+$ solvation shell is mainly composed of O atoms from FSI$^-$ and DME with minimal participation of O/F atoms from TFEO molecules. The cutoff for the Li$^+$ solvation shell is set to 2.8 Å.

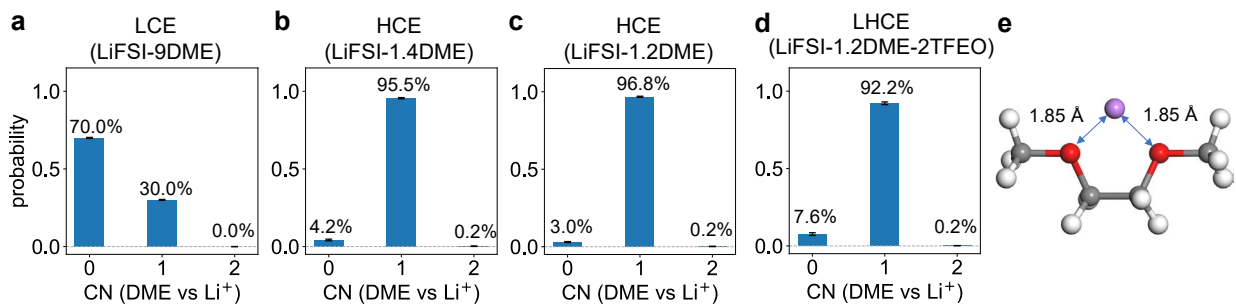

**Supplementary Fig. 2 | DME-Li$^+$ coordination.** The probability distributions of DME *vs* Li$^+$ coordination numbers (CN) in **a**, LCE (LiFSI-9DME), **b**, HCE (LiFSI-1.4DME), **c**, HCE (LiFSI-1.2DME), and **d**, LHCE (LiFSI-1.2DME-2TFEO). **e**, DFT-optimized geometry of DME molecule coordinating with one Li$^+$. It is seen that for almost all DME molecules (≥ 99.8%), each of them is coordinating with up to one Li$^+$, independent of the salt concentration. This is because when DME molecule is coordinated with one Li$^+$ through its two ether oxygen atoms there is no coordination site for a second Li$^+$. There are more DME molecules that are not coordinated with any Li$^+$ (CN = 0) in LCE compared to HCEs, which is consistent with what is shown in Fig. 3a.

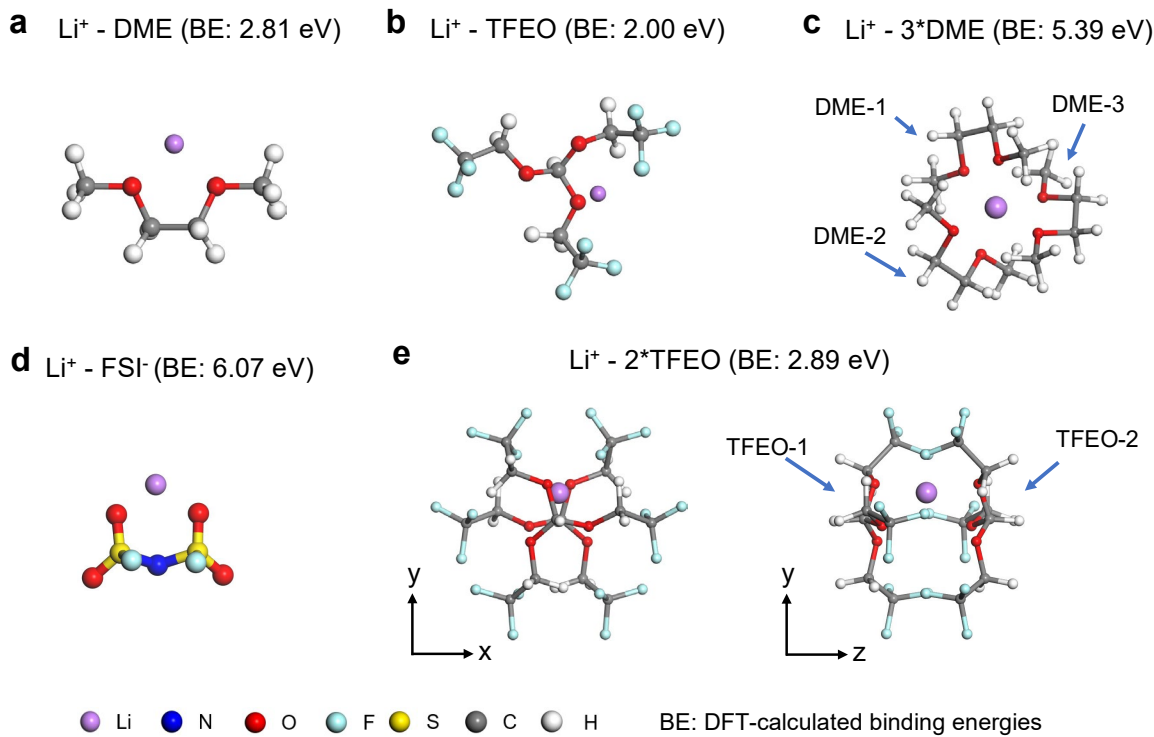

**Supplementary Fig. 3 | Binding energies (BE) of Li$^+$ to salt anion, solvent, and diluent molecules.** DFT-optimized geometry of **a**, Li$^+$ coordinating to one DME molecule, **b**, Li$^+$ coordinating with one TFEO molecule, **c**, Li$^+$ coordinating to three DME molecules, **d**, Li$^+$-FSI$^-$, and **e**, Li$^+$ coordinating to two TFEO molecules. BE for each case is given accordingly, which is defined as $E_{BE} = E_{Li^+} + E_{mol} - E_{bound}$, where $E_{Li^+}$ denotes the energy of Li$^+$, $E_{mol}$ denotes the energy of the anion (FSI$^-$), solvent (DME) cluster, or diluent (TFEO) cluster, and $E_{bound}$ denotes the energy of the bound system.

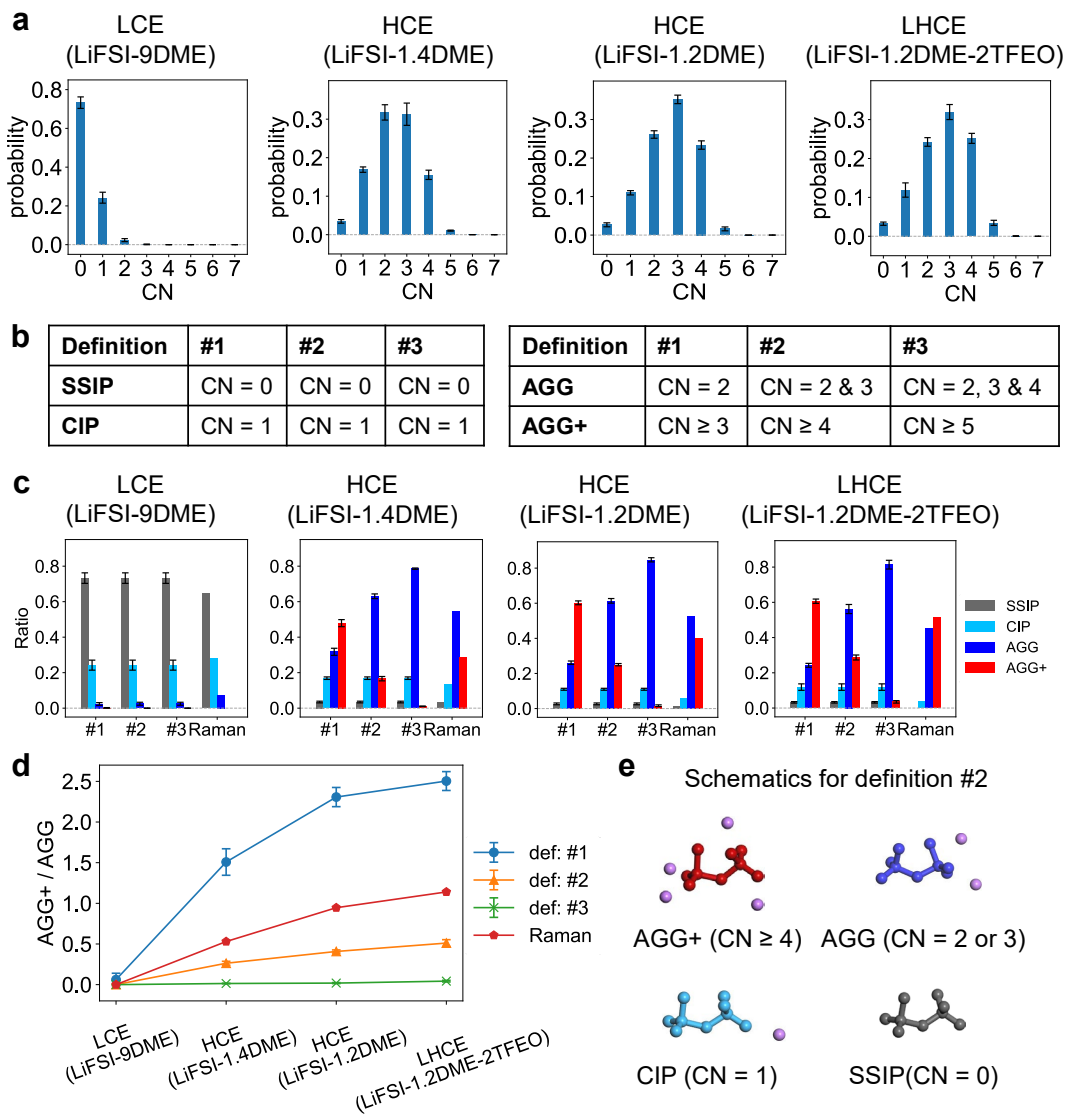

**Supplementary Fig. 4 | Definitions of salt-solvent clusters based on FSI$^-$ to Li$^+$ CNs. a**, MD-calculated probability distributions of coordination number of FSI$^-$ vs Li$^+$ pair in LCE (LiFSI-9DME), HCE (LiFSI-1.4DME), HCE (LiFSI-1.2DME) and LHCE (LiFSI-1.2DME-2TFEO), respectively, at 25 °C. **b**, Three possible definitions (#1, #2 and #3) that are used to count SSIP, CIP, AGG, and AGG+ based on MD simulations and analyses. **c**, Calculated ratios of SSIP, CIP, AGG, and AGG+ in LCE, HCEs and LHCE at 25 °C based on three different definitions and compared to results from Raman deconvolution analysis. **d**, Comparison among the ratios of AGG+/AGG calculated using different computational definitions and through Raman deconvolution analysis. It is seen that definition #2 gives the most accurate description of the salt-solvent cluster ratios when compared with the ratios calculated through Raman. Thus, definition #2 is used in this work for all analyses. **e**, Schematic and CNs for SSIP, CIP, AGG, and AGG+ using the definition #2 for the solvent-salt clusters.

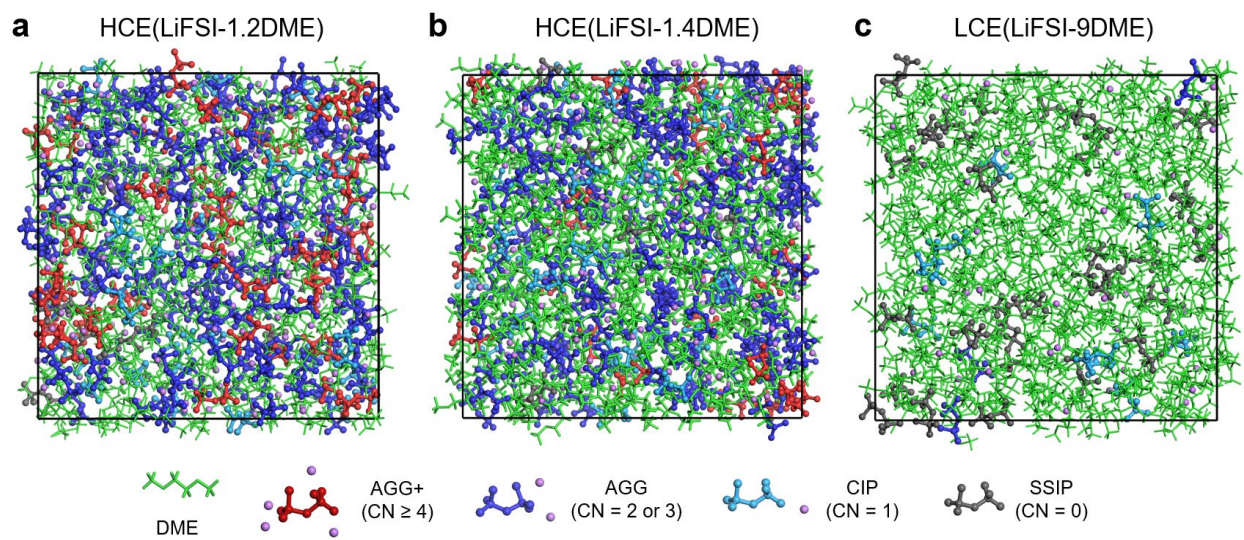

**Supplementary Fig. 5 | Salt-solvent clusters in HCEs and LCE.** Snapshots at the end of MD simulations for, **a**, HCE (LiFSI-1.2DME), **b**, HCE (LiFSI-1.4DME) and **c**, LCE (LiFSI-9DME). All LiFSI and DME molecules are shown in the snapshots. Different types of salt-solvent clusters (AGG+, AGG, CIP, and SSIP) are color-encoded for demonstration of their spatial distributions. The black lines indicate the boundaries of the simulation cells.

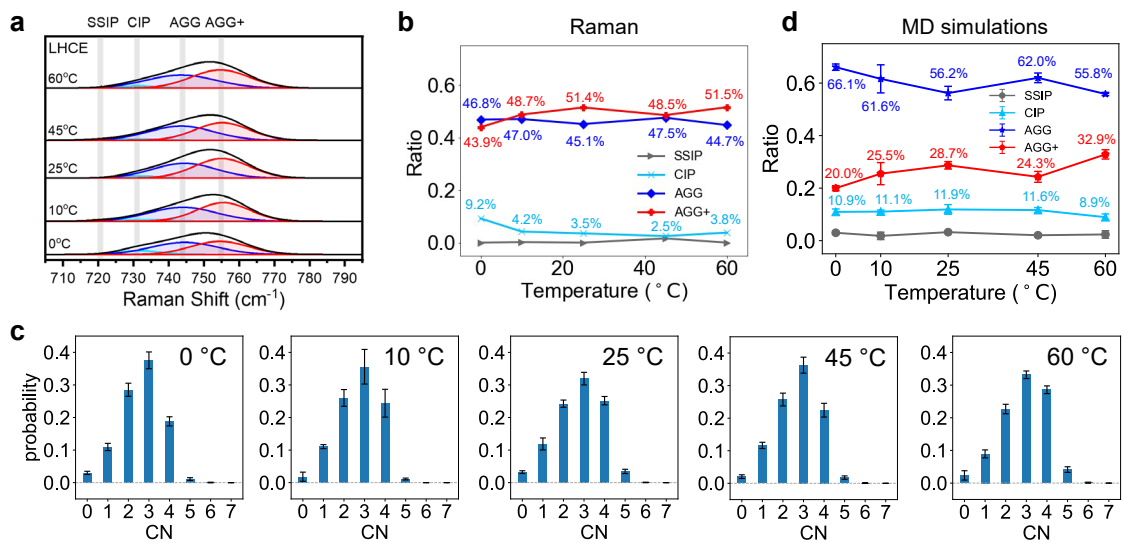

**Supplementary Fig. 6 | Salt-solvent cluster ratios as functions of temperatures in LHCE (LiFSI-1.2DME-2TFEO). a**, Raman peak deconvolution of Li$^+$-FSI$^-$ interactions and, **b**, corresponding calculated ratios of SSIP, CIP, AGG and AGG+ for LHCE (LiFSI-1.2DME-2TFEO) at different temperatures (0 °C, 10 °C, 25 °C, 45 °C, and 60 °C). **c**, Probability distributions of coordination number of FSI$^-$ *vs* Li$^+$ pair in LHCE at different temperatures calculated from MD simulations and analysis. **d**, MD-calculated probabilities of SSIP, CIP, AGG, and AGG+ at different temperatures for LHCE. Ratio percentages are included for CIP, AGG, and AGG+ of LHCE at different temperatures for both Raman and MD.

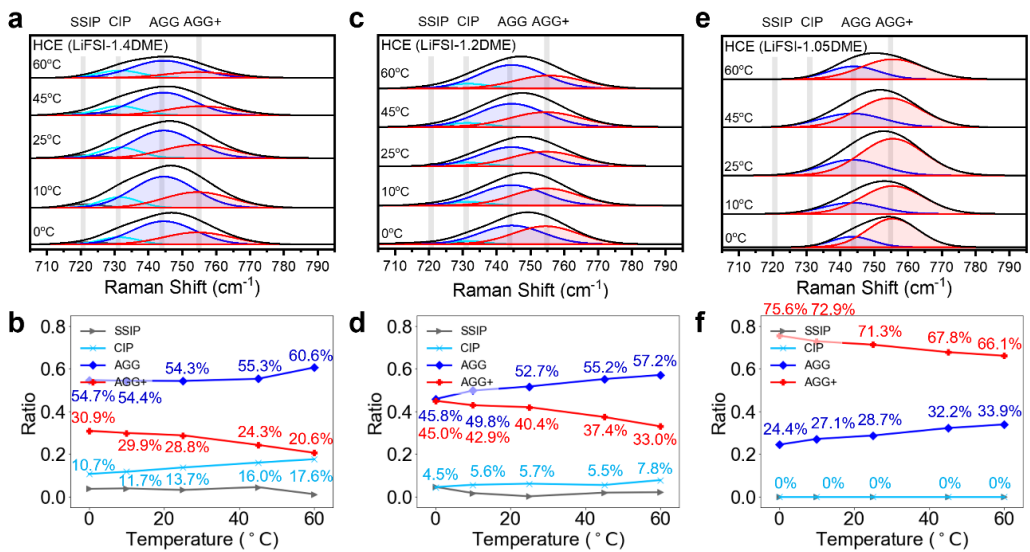

**Supplementary Fig. 7 | Salt-solvent cluster ratios as functions of temperatures in HCEs.** (Top) Raman peak deconvolution of $Li^+$-$FSI^-$ interactions and (bottom) corresponding calculated ratios of SSIP, CIP, AGG and AGG+ for **a-b**, HCE (LiFSI-1.4DME), **c-d**, HCE (LiFSI-1.2DME), and **e-f**, binary electrolyte near the solubility limit (LiFSI-1.05DME). Ratio percentages are included for all CIP, AGG, and AGG+ structures at different temperatures of each electrolyte.

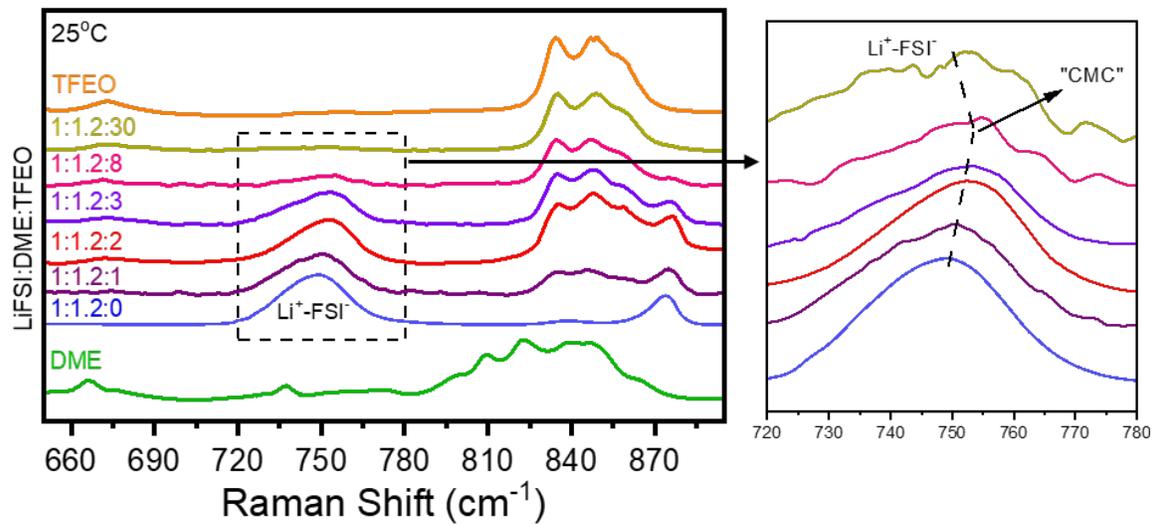

**Supplementary Fig. 8 | Raman spectra of LHCE as a function of TFEO concentration.** Raman spectra at 25°C for (bottom to top) DME (green), HCE (LiFSI-1.2DME as blue), LHCEs of LiFSI-1.2DME-$x$TFEO molar ratios ($x$ between 1 and 30), and TFEO (orange). Blue-shifting can be seen for $Li^+$-$FSI^-$ coordination peak when TFEO concentration increases up to $x=8$, which corresponds to more AGG+ in LHCE, and red-shifting can be seen for $Li^+$-$FSI^-$ coordination peak when TFEO concentration further increases, suggesting damage of micelle-like structures.

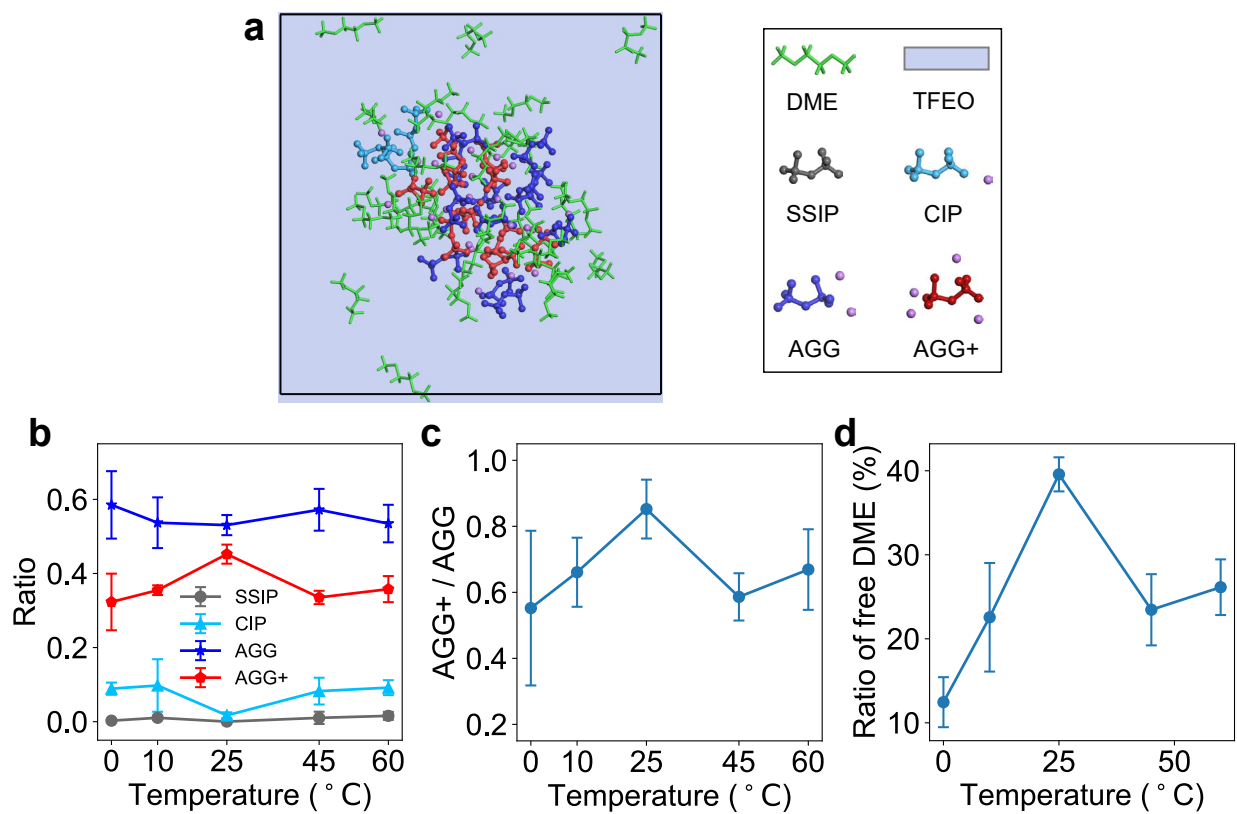

**Supplementary Fig. 9 | MD simulations of LHCE (LiFSI-1.2DME-8TFEO). a**, MD trajectory snapshot showing the spatial distributions of salt-solvent clusters in LHCE (LiFSI-1.2DME-8TFEO). The green stick model stands for DME molecule, light blue area for TFEO network, red ball-and-stick model for AGG+, blue ball-and-stick model for AGG, cyan ball-and-stick model for CIP, and dark grey ball-and-stick model for SSIP, and the square black lines indicate simulation boundaries. **b**, MD-calculated ratios of SSIP, CIP, AGG, and AGG+ as functions of temperature in LHCE (LiFSI-1.2DME-8TFEO). **c**, MD-calculated AGG+/AGG ratio as function of temperature in LHCE (LiFSI-1.2DME-8TFEO). **d**, Ratio of free DME molecules (not coordinating with any Li$^+$) as a function of temperature in LHCE (LiFSI-1.2DME-8TFEO).

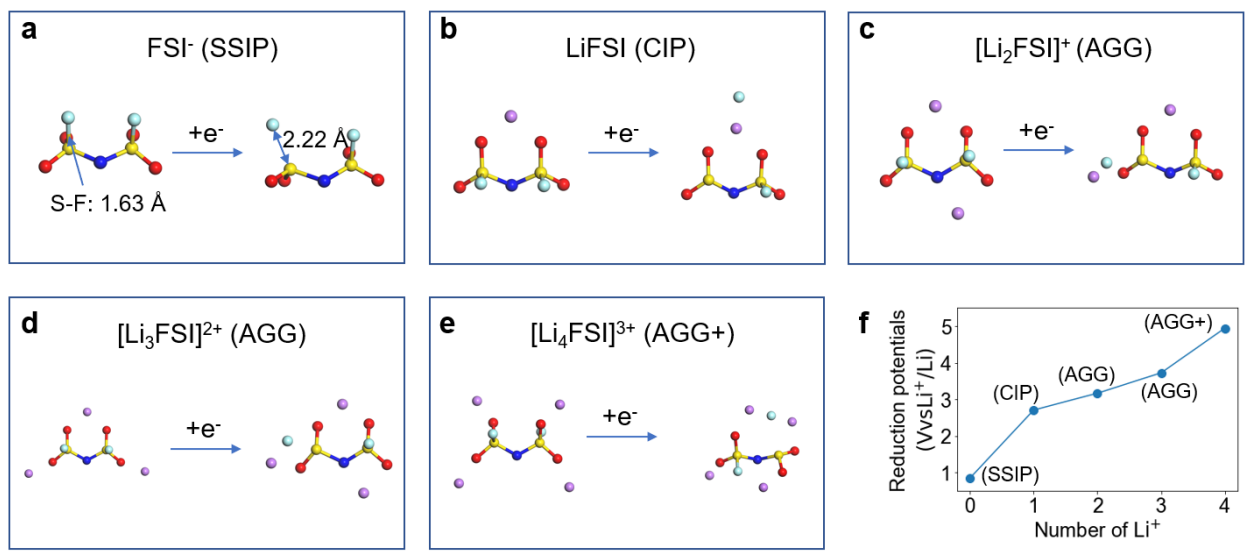

**Supplementary Fig. 10 | DFT-calculated reduction potentials for different salt clusters.** Reduction reactions and corresponding reduction potentials (*vs* $Li^+/Li$) for **a**, $FSI^-$, **b**, LiFSI, **c**, $[Li_2FSI]^+$, **d**, $[Li_3FSI]^{2+}$, and **e**, $[Li_4FSI]^{3+}$. The aggregate types (SSIP, CIP, AGG, and AGG+) are shown in parentheses accordingly. **f**, Reduction potentials as functions of the number of $Li^+$ ions that are coordinated with one $FSI^-$.

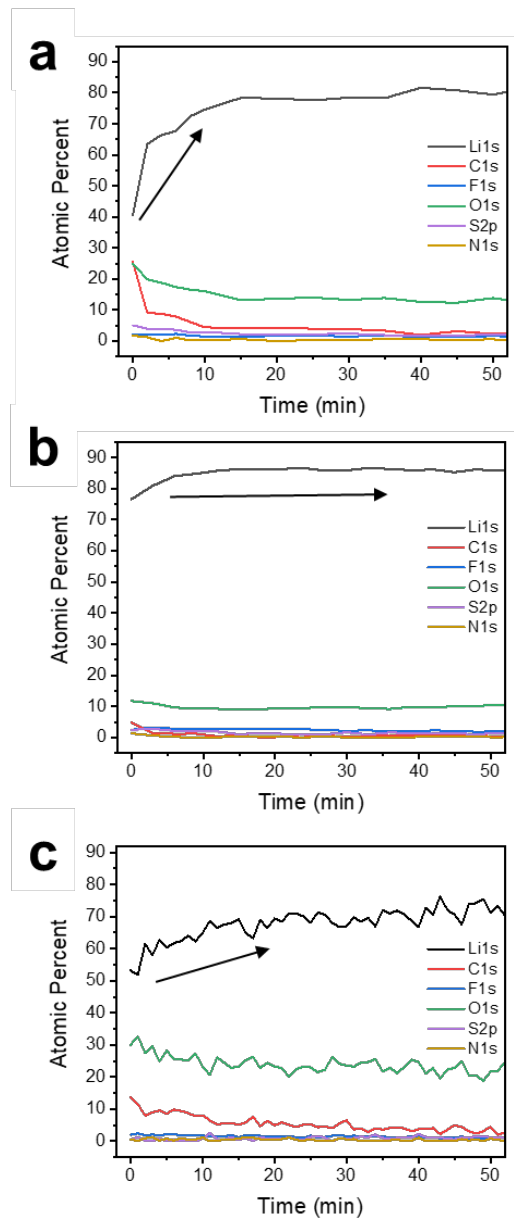

**Supplementary Fig. 11 | Elemental depth profiling.** XPS depth profiles of stripped (i.e., delithiated) Li foils after formation cycles at **a**, 10 °C, **b**, 25 °C, and **c**, 45 °C. At 25 °C, a higher Li atomic concentration and brief change in its concentration over time of sputtering reveals its thin thickness in comparison to the decomposition surface seen at 10 °C and 45 °C. At 10 °C, the change in Li concentration changes rapidly over the first 10-15 minutes of sputtering, followed by relatively steady composition after further sputter. At 45 °C, a slow and steady change in composition is seen over the extent of sputtering, revealing either thick or highly non-uniform decomposition surface.

**Supplementary Table 1 | SEI comparison of elements from XPS.** XPS depth profile atomic ratios of fluorine, sulfur, and nitrogen in comparison to carbon, averaged over the range of the SEI for each electrolyte and temperature (a cut-off of 20 minutes sputtering was used for depth profiles with linear trends, 10 °C and 45 °C). The higher ratio of F/C, S/C and N/C ratio at 25 °C suggests more inorganic and less organic material in the decomposition surface.

|  | 10°C | 25°C | 45°C |
|---|---|---|---|
| F:C Ratio | 0.25 | 1.07 | 0.24 |
| S:C Ratio | 0.45 | 1.86 | 0.097 |
| N:C Ratio | 0.074 | 1.63 | 0.086 |

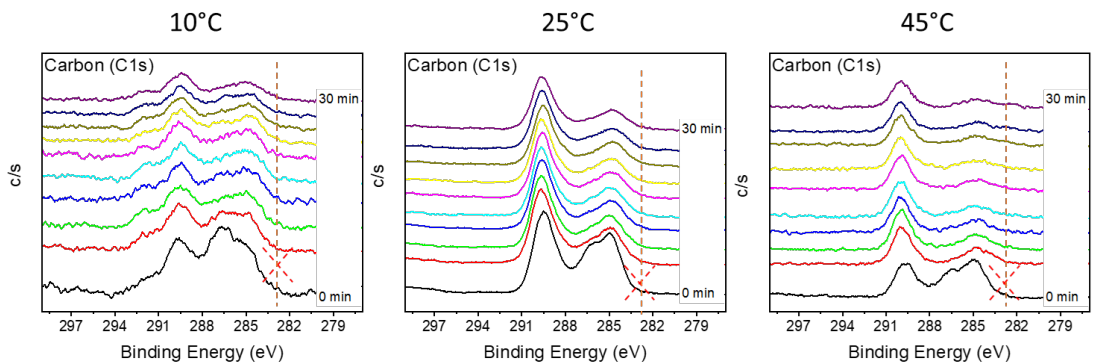

**Supplementary Fig. 12 | Carbon peak depth profiling of HCE (LiFSI-1.2DME) at different temperatures.** XPS depth profiles of stripped (i.e., delithiated) Li foils after formation cycles for carbon at 10 °C, 25 °C, and 45 °C. Obvious C-Li bonds (Fig. 5d) and C-F bonds (Fig. 5f) in SEI formed only at 10 ºC were found. Moreover, C-Li bonds almost do not exist in HCE without TFEO. Those suggest Li-C formation is due to the decomposition of TFEO and more organic components formed in the SEI at 10 ºC would permit further decomposition mechanisms of the diluent or solvent, resulting in obvious C-F and C-Li peaks.

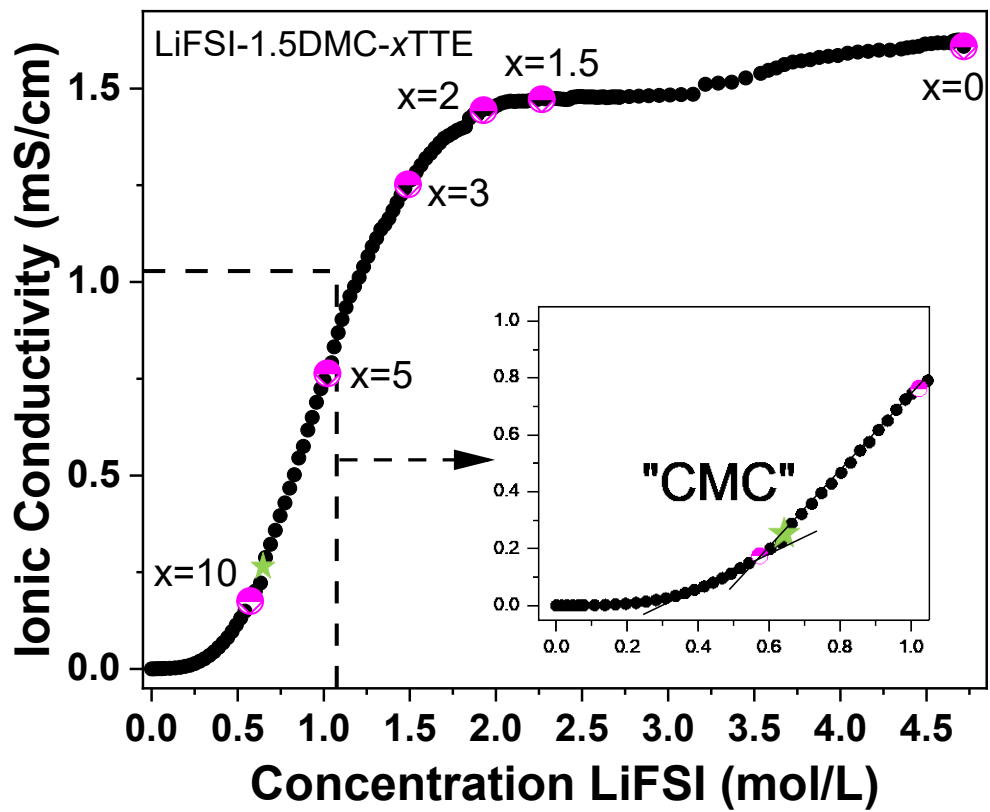

**Supplementary Fig. 13 | Ionic conductivity of LiFSI-DMC-TTE electrolyte.** Ionic conductivities of LiFSI-1.5DMC-$x$TTE solutions as a function of LiFSI salt concentration, acquired between 21-22 °C. Specific solutions (pink circles), as well as the "CMC" (in the inset as a green star) are noted.

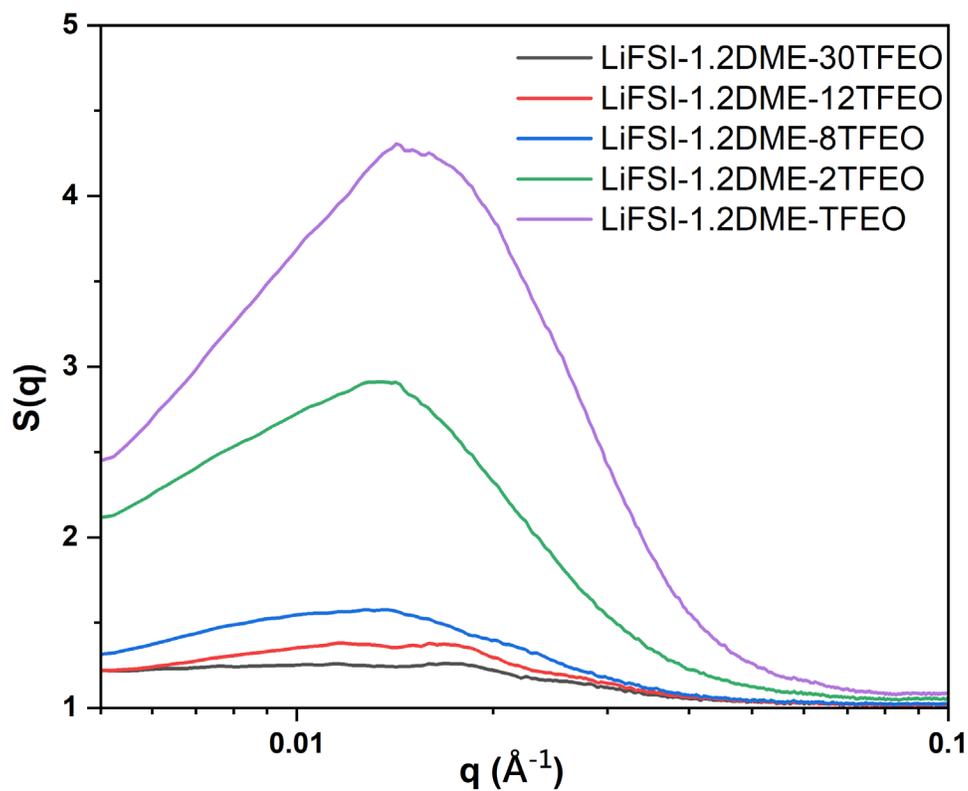

**Supplementary Fig. 14 | Structure factors, S(q), of LiFSI-1.2DME-*x*TFEO electrolytes.** Structure factor derived from SAXS patterns with TFEO in different molar ratios in LHCEs (x = 1, 2, 8, 12 and 30). Peak intensity increases when diluent concentration decreases, suggesting more micelle-like structures are interacting with each other.

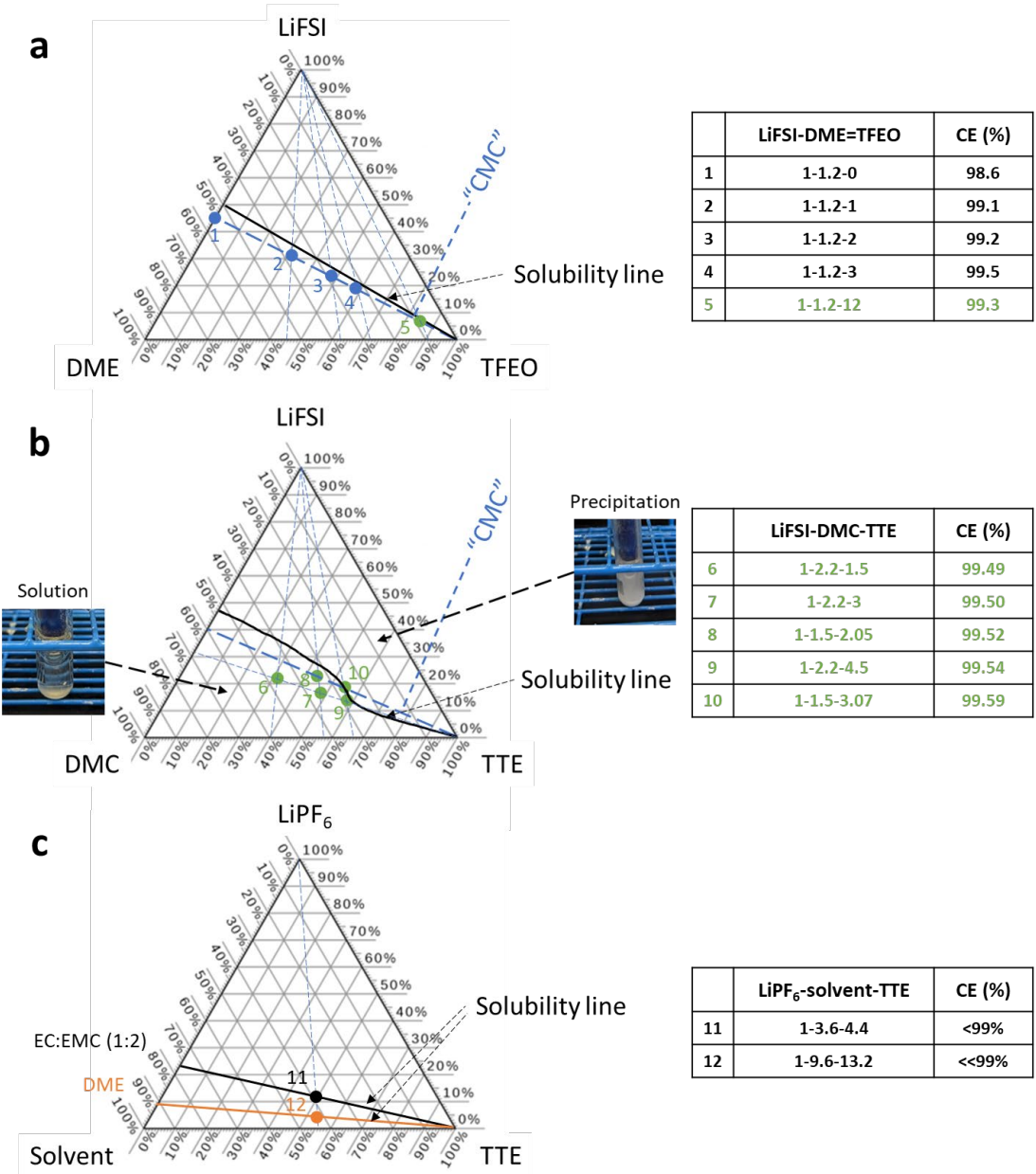

**Supplementary Fig. 15 | Ternary phase diagrams of different LHCE systems. a**, LiFSI-DME-TFEO.[1] **b**, LiFSI-DMC-TTE. **c**, LiPF$_6$-DME-TTE and LiPF$_6$-solvent-TTE (solvent as DME in orange, EC-EMC (1:2 by mole) in black).[2] The electrolyte formulation (molar ratio) and corresponding CE values are shown in the tables. Electrolyte formulations highlighted in green were tested in this work, utilizing the modified Aurbach (i.e., "Method 3"), referred elsewhere.[3] All other CE values are from literature, using similar methods. In the LiFSI-DMC-TTE system, the solubility line was approximated by experiments here and in literature,[4] as exemplified with the solution and precipitation images.

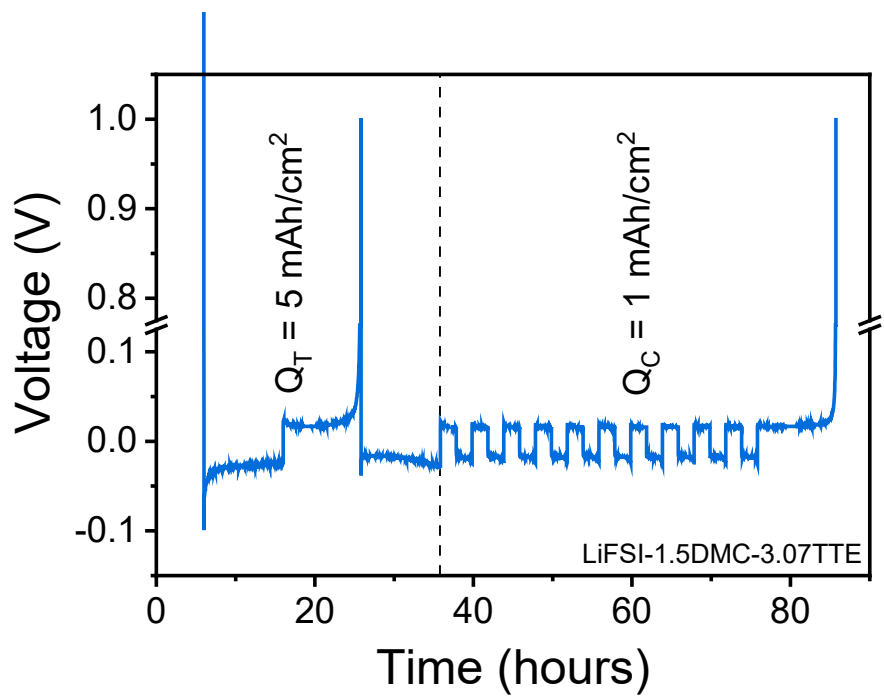

**Supplementary Fig. 16 | Coulombic efficiency (CE) voltage profile of LiFSI-1.5DMC-3.07TTE LHCE.** Coulombic efficiency averaged 99.59% over the 10 cycles. $Q_T$ refers to capacity during formation cycles, while $Q_C$ refers to capacity during CE-measured cycles. This CE testing design is based on the modified Aurbach (i.e., "Method 3"), referred elsewhere.[3]

**Supplementary Table 2 | CE values of different LHCEs in literature.** All CE values are accomplished in Li||Cu cells. Capacity, current, and number of cycles ran are included. Generally, CE would increase with increasing cycle numbers due to the formation of initial SEIs on Cu surfaces.

| Electrolyte | Capacity (mAh/cm$^2$) | Current (mA/cm$^2$) | Cycles Ran | CE | Ref. |
|---|---|---|---|---|---|
| LiFSI-DMC-TTE (1-1.5-1.5 by mole) | 1 | 0.2 | 50 | 99.05 | 4 |
| LiFSI-DME-BTFE (1-1.2-3 by mole) | 1 | 0.5 | 10 | 99.4 | 5 |
| LiFSI-DME-BTFEC (1-1.2-3 by mole) | 1 | 0.5 | 10 | 96.8 | 5 |
| LiFSI-DME-TFEB (1-1.4-3 by mole) | 1 | 0.5 | 10 | 95.4 | 5 |
| LiFSI-DME-TFEO (1-1.2-3 by mole) | 1 | 0.5 | 10 | 99.5 | 5,6 |
| LiFSI-DME-TTE (1-1.2-3 by mole) | 1 | 0.5 | 10 | 99.5 | 5 |
| LiFSI-DME-TTE (1-1.2-3 by mole) | 1 | 0.5 | 300 | 99.3 | 7 |
| LiTFSI-DME-DOL-TME (1-1.5-1.5-6 by mole) | 1 | 0.5 | 10 | 99.13 | 8 |
| 2.5M LiFSI in DMC-BTFE (1-1 by mole) | 1 | 0.5 | 10 | 99.5 | 9 |
| 0.4M LiTFSI + 0.4M LiNO3 + 0.1M LiHFDF in DME-DOL-TFMTMS (48-17-35 by vol) | 1 | 0.5 | 100 | 98.5 | 10 |
| 1M LiFSI in OFE-DME (1-5 by mole) | 1 | 0.5 | 250 | 99.3 | 11 |
| 1M LiFSI in isoxazole-TTE (1-4.5 by vol) + 10 wt.% FEC | 1 | 1 | 10 | 98.6 | 12 |
| LiFSI-DME-FEC-OFE (1-1-1.4-3 by mole) | 2 | 1 | 200 | 98.9 | 13 |
| 1.2M LiFSI in TEP-BTFE (1-3 by mole) | 1 | 0.5 | 10 | 99.2 | 14 |
| LiFSI-TMS-TTE (1-3-3 by mole) | 1 | 0.5 | 150 | 98.8 | 15 |

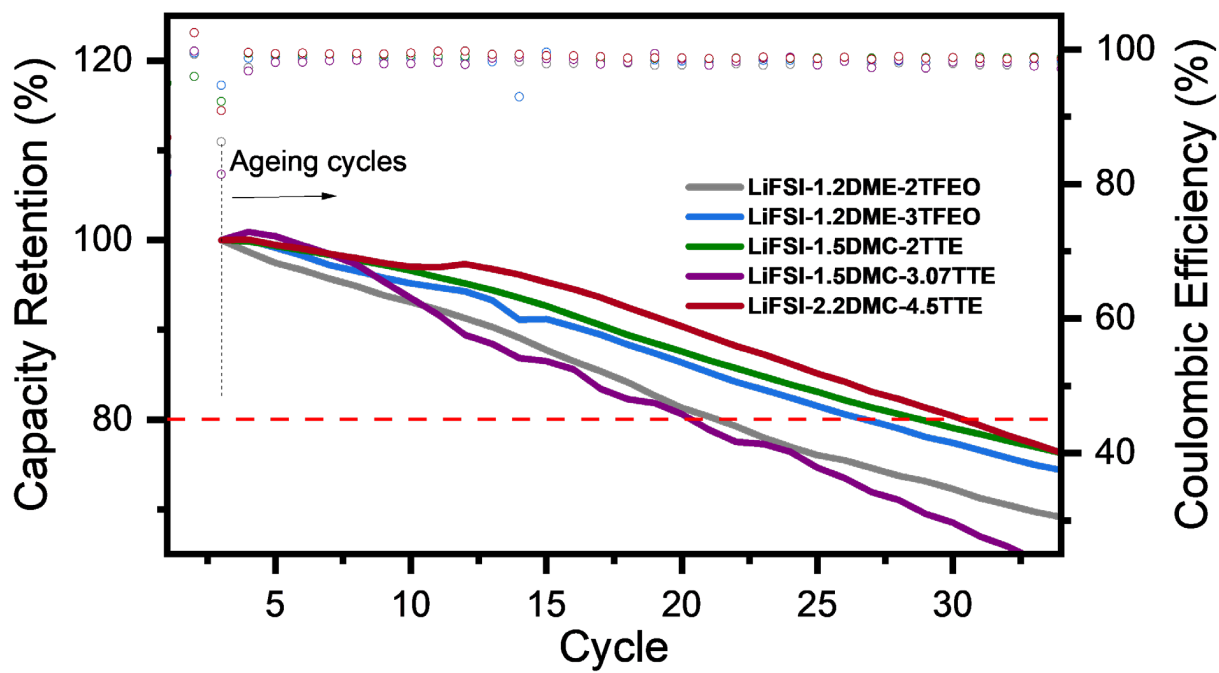

**Supplementary Fig. 17 | Cu-NMC811 coin cell cycling performance for elected LHCEs.** Coulombic efficiency and discharge capacity normalized from the first aging cycle show for the best performing LHCEs from CE tests (Supplementary Fig. 16). Cycling protocol followed the same process as Li-NMC811 coin cell tests. Dashed lines denote cycle in which the capacity retention falls below 80%.

# MD Simulation Details

## 1. Force field and its validation

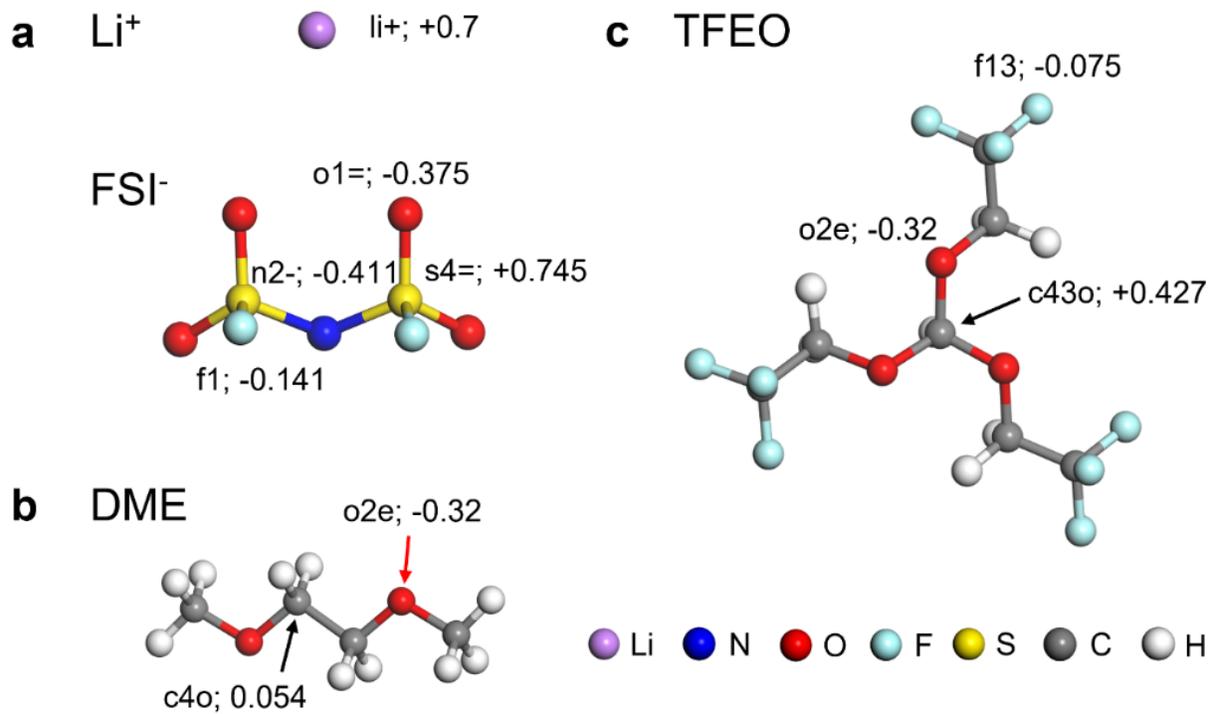

**Supplementary Fig. 18 | Atom types and atomic charges used in MD simulations.** Representative atom types and atomic charges in **a**, LiFSI, **b**, DME, and **c**, TFEO used in MD simulations. Note that all atomic charges in LiFSI are scaled by 0.7 to account for the ion-ion and ion-dipole interactions.

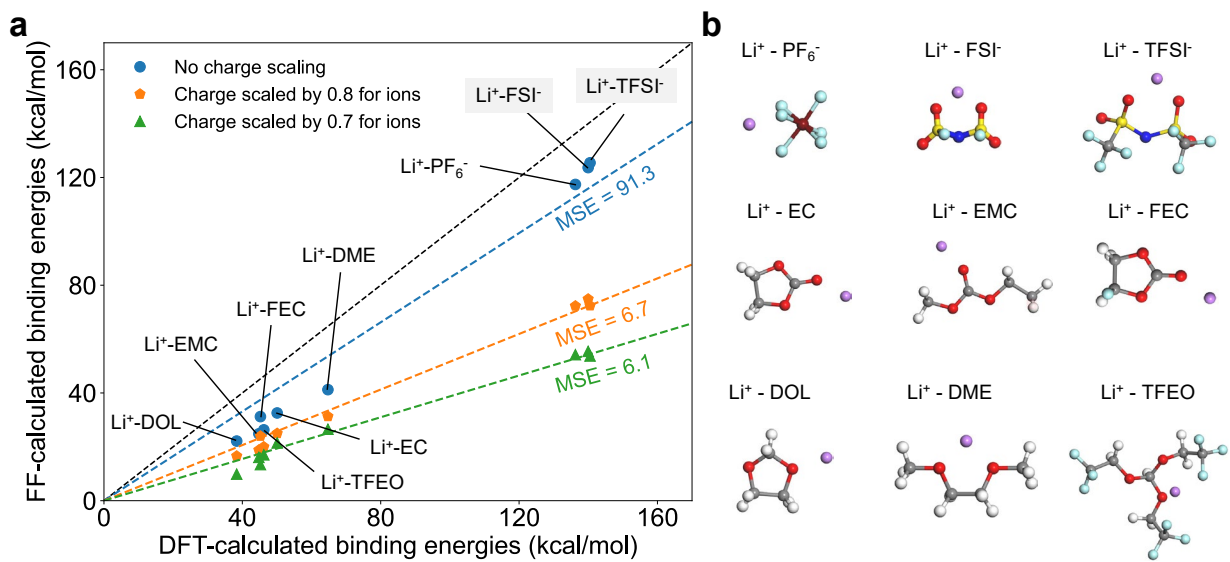

**Supplementary Fig. 19 | DFT- *vs* force field (FF)-calculated binding energies. a**, Binding energies calculated based on COMPASS III force field (FF) as functions of binding energies calculated based on DFT method. Binding energy (BE) is defined as $E_{BE} = E_{Li^+} + E_{mol} - E_{bound}$, where $E_{Li^+}$ denotes the energy of Li$^+$, $E_{mol}$ denotes the energy of the molecular species (salt anion, solvent or diluent), and $E_{bound}$ denotes the energy of the bound system. Blue dots refer to the FF-calculated binding energies without charge scales, while orange and green dots refer to the binding energies calculated based on charge scales of 0.8 and 0.7, respectively. The blue, orange, and green dash lines correspond to linear fittings with intercept of zero. It is seen that the mean squared errors (MSE, in kcal/mol) of the linear fittings are largely reduced with charge scales of 0.8 and 0.7, which indicate the cation-anion and cation-solvent interactions become more balanced. **b**, atomic structures of Li$^+$-coordinated species. Gray, white, red, light blue, purple, brown, yellow, and dark blue atoms represent C, H, O, F, Li, P, S, and N atoms, respectively.

Supplementary Tables 3, 4 and 5 list the molar ratio, number of salt and solvent species in the simulation cell, scaled charge, temperature, simulation cell size, salt concentration, density, Li$^+$ diffusivity, Li$^+$ conductivity, and Li$^+$ coordination numbers with O/F atoms in salt anions and solvents. For calculations of coordination numbers, we used a cutoff of 2.8 Å. Note that the blue texts in brackets are experimentally measured data and the red texts in brackets are MD-simulated data with APPLE&P force field by Oleg Borodin and Grant D. Smith.[16] In Supplementary Table 3 and Table 4, both the experimental and simulation data are taken from Qian et al.[17] The experimental data was obtained at 25 °C and the MD simulations with APPLE&P force field were conducted at 60 °C. In Supplementary Table 5, the experimentally measured data (for the 1 M LiPF$_6$ in EC-EMC-DEC (2-2-1 by mole) electrolyte) and the APPLE&P force field simulated data is from Hayashi et al.[18] and Borodin and Smith,[16] both of which were obtained at 25 °C.

**Supplementary Table 3 | MD data of LCE (1 M LiFSI in DME).**

| Simulated electrolyte | | 1M LiFSI-DME | | | | | |
|---|---|---|---|---|---|---|---|
| Molar ratio (DME:LiFSI) | | 9:1 (9:1) (8.9:1) | | | | | |
| # of DME and LiFSI in simulation cell | | 360 DME and 40 LiFSI (576 DME and 64 LiFSI) | | | | | |
| **Charge scale of LiFSI** | | **0.8** | | | **0.7** | | |
| Temperature (ºC) | | -20 | 20 | 60 | -20 | 20 | 60 |
| Simulation cell size (Å) | | 39.4 | 39.9 | 40.5 | 39.4 | 39.9 | 40.5 |
| Concentration (M) | | 1.09 | 1.05 | 1.00 (0.96) | 1.09 | 1.04 | 1.00 (0.96) |
| Density (g/cm$^3$) | | 1.09 | 1.04 | 1.00 (0.96) | 1.08 | 1.04 | 1.00 (0.96) |
| Li$^+$ diffusivity (×10$^{-10}$ m$^2$/s) | | 0.66 | 2.29 | 5.04 | 1.21 | 3.99 | 7.60 |
| Li$^+$ conductivity (mS/cm) | | 3.20 | 9.14 (16.9) | 17.0 (23.2) | 5.81 | **15.9** (16.9) | **25.6** (23.2) |
| coordination number (cutoff: 2.8 Å) | total (Li-O/F) | 5.75 | 5.52 | 5.23 | 5.63 | 5.31 | 5.03 |
| | Li-O(DME) | 4.77 | 4.29 | 3.70 (4.42) | 5.52 | 4.93 | **4.50** (4.42) |
| | Li-O(FSI) | 0.98 | 1.22 | 1.52 (0.42) | 0.11 | 0.37 | **0.52** (0.42) |
| | Li-F | 0.00 | 0.01 | 0.01 | 0.00 | 0.01 | 0.01 |

**Supplementary Table 4 | MD data of HCE (4 M LiFSI in DME).**

| Simulated electrolyte | 4M LiFSI-DME | | | | | |
|---|---|---|---|---|---|---|
| Molar ratio (DME:LiFSI) | 1.4:1 (1.4:1) (1.4:1) | | | | | |
| # of DME and LiFSI in simulation cell | 288 DME and 200 LiFSI (488 DME and 320 LiFSI) | | | | | |
| **charge scale of LiFSI** | 0.8 | | | 0.7 | | |
| Temperature (ºC) | -20 | 20 | 60 | -20 | 20 | 60 |
| Simulation cell size (Å) | 40.8 | 41.1 | 41.4 | 41.0 | 41.3 | 41.7 |
| Concentration (M) | 4.88 | 4.78 | **4.68** (4.24) | 4.82 | 4.72 | **4.59** (4.24) |
| density (g/cm$^3$) | 1.53 | 1.50 | 1.47 (1.33) | 1.53 | 1.48 | 1.47 (1.33) |
| Li$^+$ diffusivity (×10$^{-10}$ m$^2$/s) | 0.003 | 0.018 | 0.053 | 0.01 | 0.047 | 0.38 |
| Li$^+$ conductivity (mS/cm) | 0.072 | 0.33 | **0.84** (5.7) (4.2) | 0.21 | 0.84 (5.7) | **5.82** (4.2) |
| coordination number (cutoff: 2.8 Å) — total (Li-O/F) | 5.35 | 5.20 | 5.11 | 5.18 | 5.07 | 4.99 |
| Li-O(DME) | 2.17 | 2.15 | 2.07 (2.60) | 2.14 | 2.49 | 2.02 (2.60) |
| Li-O(FSI) | 3.12 | 3.00 | 2.98 (1.92) | 2.98 | 2.51 | 2.91 (1.92) |
| Li-F | 0.06 | 0.05 | 0.06 | 0.06 | 0.07 | 0.06 |

**Supplementary Tables 5 | MD data of 1M LiPF$_6$ in EC-EMC (3-7 by vol).**

| Simulated electrolyte | 1M LiPF$_6$-EC:EMC(3:7, v:v) | | | | | |
|---|---|---|---|---|---|---|
| Molar ratio (EC:EMC:LiPF$_6$) | 13:20:3 | | | | | |
| # of EC, EMC and LiPF$_6$ | 156 EC, 240 EMC and 36 LiPF$_6$ | | | | | |
| **charge scale of LiPF$_6$** | **0.8** | | | **0.7** | | |
| Temperature (°C) | -20 | 20 | 60 | -20 | 20 | 60 |
| Simulation cell size (Å) | 39.0 | 39.5 | 40.0 | 39.1 | 39.6 | 40.1 |
| Concentration (M) | 1.01 | 0.97 | 0.93 | 1.00 | 0.96 | 0.93 |
| Density (g/cm$^3$) | 1.23 | 1.19 | 1.14 | 1.22 | 1.18 | 1.14 |
| Li$^+$ diffusivity (×10$^{-10}$ m$^2$/s) | 0.70 | 2.52 (~3.5) | 4.98 | 1.06 | 3.48 (~3.5) | 9.04 |
| Li$^+$ conductivity (mS/cm) | 3.10 | **9.32** (~10) (~11) | 15.6 | 4.68 | **12.8** (~10) (~11) | 28.2 |
| coordination number (cutoff: 2.8 Å) total (Li-O/F) | 5.01 | 4.91 | 4.85 | 4.61 | 4.54 | 4.48 |
| Li-O(EC) | 2.58 | 2.35 | 2.27 | 2.48 | 2.29 | 2.18 |
| Li-O(EMC) | 1.82 | 1.74 | 1.52 | 1.86 | 1.78 | 1.67 |
| Li-F | 0.61 | 0.82 | 1.06 | 0.27 | 0.47 | 0.63 |

**Supplementary Table 6 | MD-calculated ratios of SSIP, CIP, AGG and AGG+ that are obtained using scaled charges of 0.7 and 0.8, as well as through Raman deconvolution analysis.** It is seen that scaled charge of 0.7 gives reasonable description of ion pairing and aggregation for both LCE and HCE when compared to Raman data. The SSIP ratio is significantly underestimated with a scaled charge of 0.8 in LCE when compared to Raman data.

| Electrolyte | Method | SSIP | CIP | AGG | AGG+ |
|---|---|---|---|---|---|
| LCE (LiFSI-9DME) | MD (Scale: 0.7) | 0.733 | 0.242 | 0.025 | 0.000 |
| | MD (Scale: 0.8) | 0.185 | 0.376 | 0.424 | 0.015 |
| | Raman | 0.646 | 0.279 | 0.075 | 0.000 |
| HCE (LiFSI-1.4DME) | MD (Scale: 0.7) | 0.024 | 0.131 | 0.629 | 0.216 |
| | MD (Scale: 0.8) | 0.010 | 0.065 | 0.621 | 0.304 |
| | Raman | 0.032 | 0.137 | 0.543 | 0.288 |

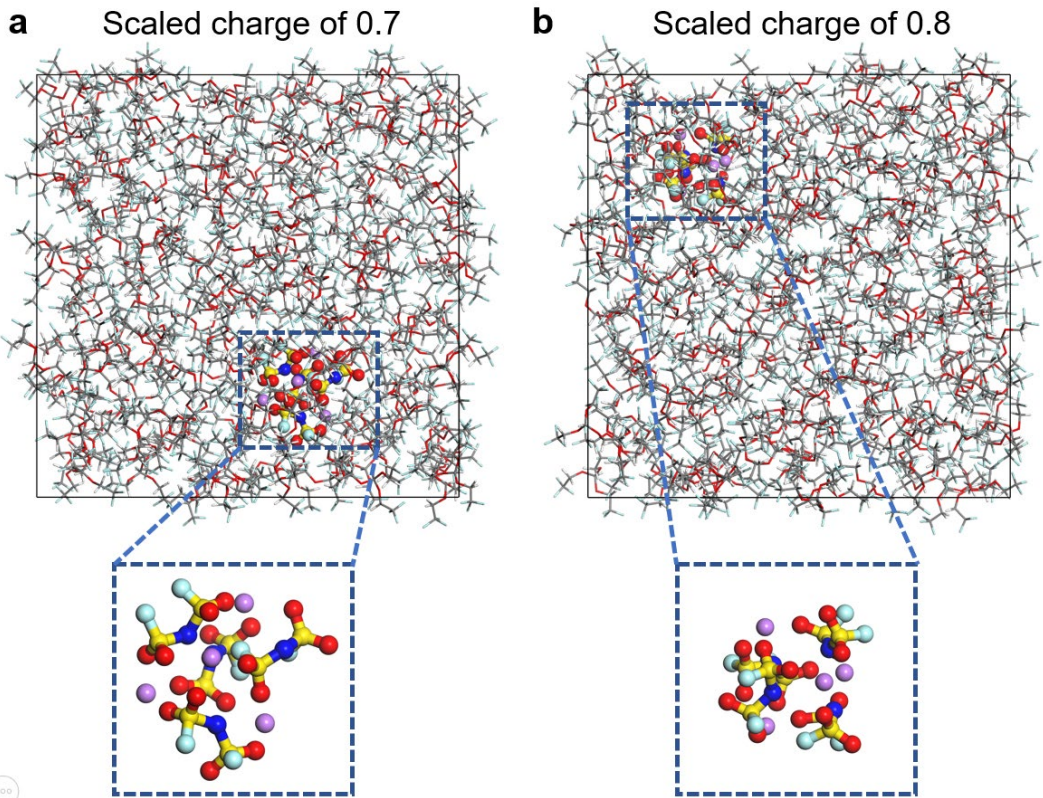

**Supplementary Fig. 20 | MD-simulated structure of 4 LiFSI molecules in TFEO matrix (with 200 TFEO molecules) for different scaled charges.** Scaled charge of **a,** 0.7 and **b,** 0.8, where in both simulations the cations and anions were initially uniformly separated in the TFEO diluent. By the end of the simulation (22.0 ns under NPT ensemble at room temperature, 25 °C), the anions and cations formed salt clusters (with 4 $Li^+$ and 4 $FSI^-$). Both simulations reveal no solvation of LiFSI salt in TFEO, suggesting that the scaled charge of 0.7 or 0.8 makes no difference in balancing the ion charges and partial charges with TFEO diluent.

## 2. Simulation Systems

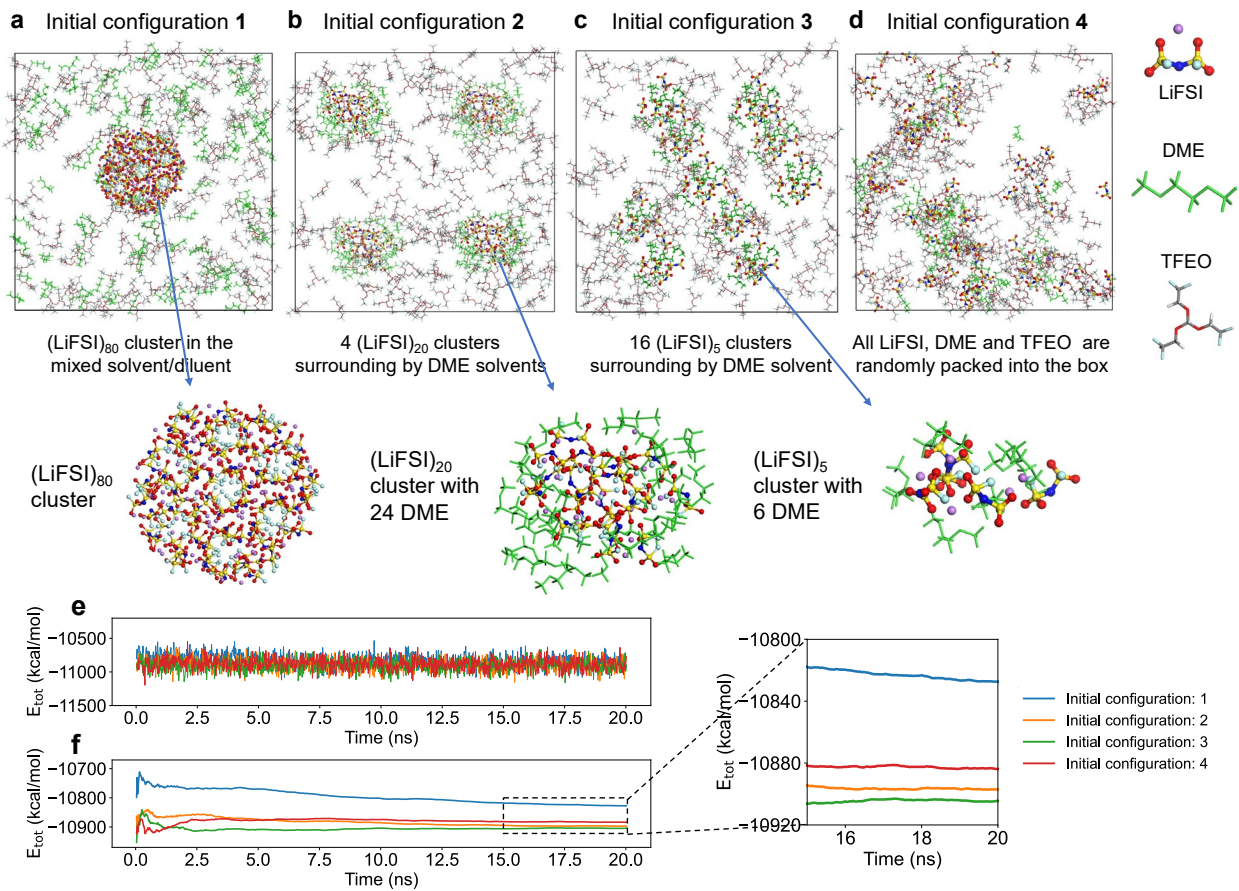

**Supplementary Fig. 21 | MD simulations of LHCE starting from several different initial configurations**. **a**, Initial configuration 1 for MD simulations starting from immersing LiFSI cluster (80 LiFSI molecules, i.e., $(LiFSI)_{80}$) into the mixed DME solvent and TFEO diluent. **b**, Initial configuration 2 for MD simulations starting with 4 $(LiFSI)_{20}$ clusters surrounding by DME solvents and then TFEO diluent. **c**, Initial configuration 3 for MD simulations starting with 16 $(LiFSI)_5$ clusters surrounding by DME solvents and then TFEO diluent. **d**, Initial configuration 4 for MD simulations starting with all LiFSI, DME and TFEO being randomly packed into the simulation box. All MD simulations were conducted with same number of species (80 LiFSI, 96 DME and 160 TFEO) and same initial simulation box size (80×80×80 Å$^3$) at the same temperature (25 °C). **e**, Evolutions of total energies and **f**, time-averaged total energies of MD simulations with different initial configurations (1-4). It is seen that the MD simulation starting with initial configuration 3 gives the lowest total energy, and thus adopted for MD simulations at other temperatures as well.

**Supplementary Table 7 | Table list of number of LiFSI, DME, and/or TFEO in the simulation cells of different systems for MD simulations.**

| System | LiFSI | DME | TFEO |
|---|---|---|---|
| LHCE (LiFSI-1.2DME-2TFEO) | 80 | 96 | 160 |
| HCE (LiFSI-1.2DME) | 200 | 240 | - |
| HCE (LiFSI-1.4DME) | 200 | 280 | - |
| LCE (LIFSI-9DME) | 40 | 360 | - |
| DME-TFEO (DME-2TFEO) | - | 96 | 160 |
| LiFSI-TFEO (LiFSI:TFEO → 0) | 4 | - | 200 |
| LiFSI crystal | 432 | - | - |

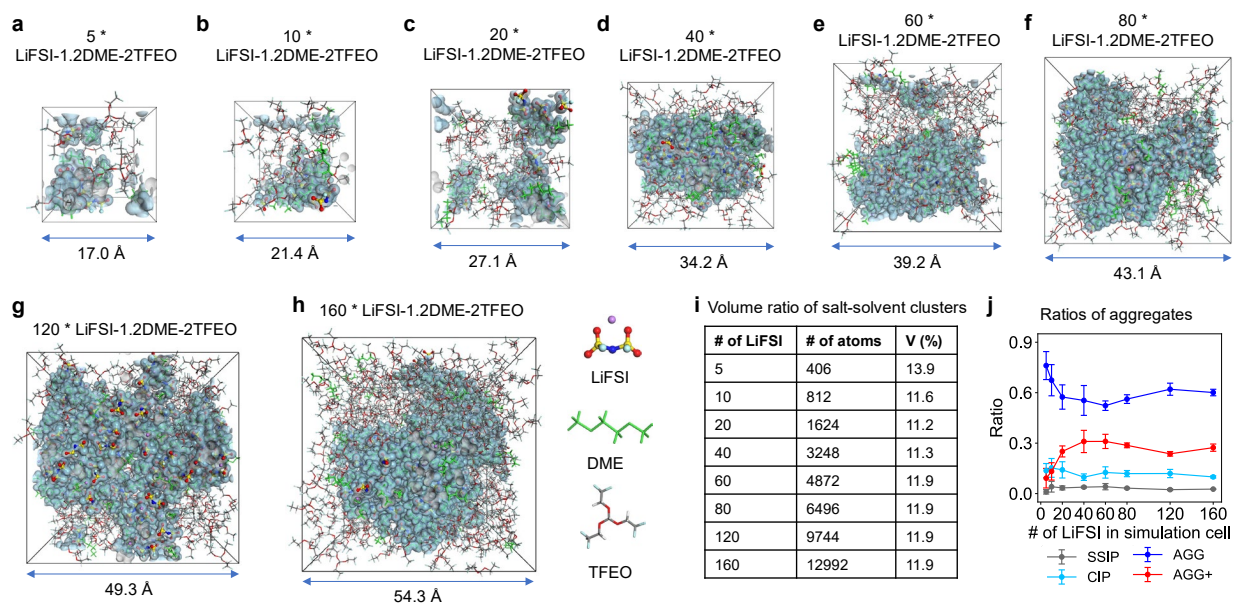

**Supplementary Fig. 22 | Perspective views of MD simulations using different simulation cells at room temperature for LHCE (LiFSI-1.2DME-2TFEO)**. **a**, 5 * LiFSI-1.2DME-2TFEO with a box size of 17.0 Å on each side; **b**, 10 * LiFSI-1.2DME-2TFEO with a box size of 21.4 Å; **c**, 20 * LiFSI-1.2DME-2TFEO with a box size of 27.1 Å; **d**, 40 * LiFSI-1.2DME-2TFEO with a box size of 34.2 Å; **e**, 60 * LiFSI-1.2DME-2TFEO with a box size of 39.2 Å; **f**, 80 * LiFSI-1.2DME-2TFEO with a box size of 43.1 Å; **g**, 120 * LiFSI-1.2DME-2TFEO with a box size of 49.3 Å; **h**, 160 * LiFSI-1.2DME-2TFEO with a box size of 54.3 Å. The vdW surfaces are shown accordingly. The salt-solvent clusters are highlighted through drawing their vdW surfaces. **i**, Number of LiFSI and atoms, as well as volume ratios of the salt-solvent clusters for LHCE with different simulation cells. **j**, Ratios of aggregates (SSIP, CIP, AGG and AGG+) for LHCE with different simulation cells.

## 3. Statistics of MD results

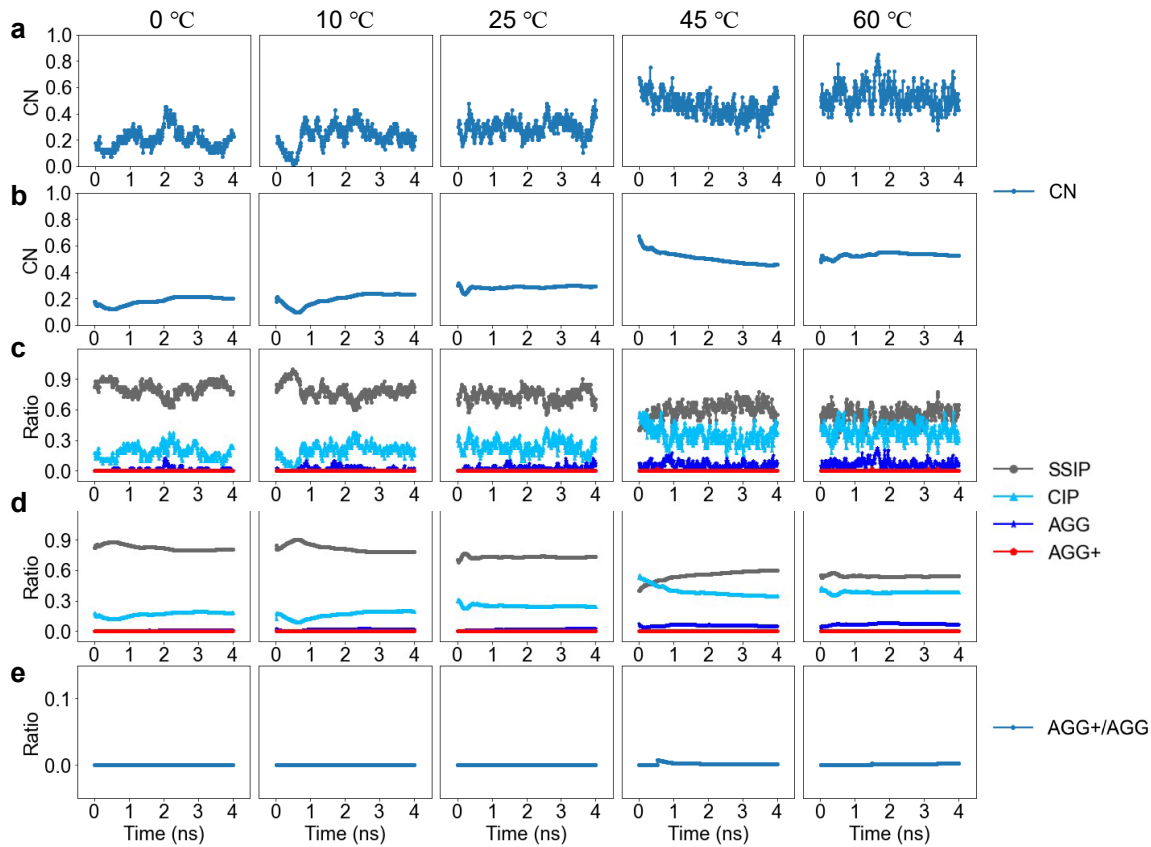

**Supplementary Fig. 23 | Statistics of coordination numbers (CN) and aggregation ratios for LCE (LiFSI-9DME).** **a**, MD-calculated FSI$^-$ *vs* Li$^+$ CN as function of time during production runs (the last 4 ns simulations). **b**, Time-averaged FSI$^-$ *vs* Li$^+$ CN as function of time. **c**, MD-calculated aggregation ratios (SSIP, CIP, AGG and AGG+) as function of time. **d**, Time-averaged aggregation ratios (SSIP, CIP, AGG and AGG+) as function of time. **e**, AGG+/AGG ratio as function of time.

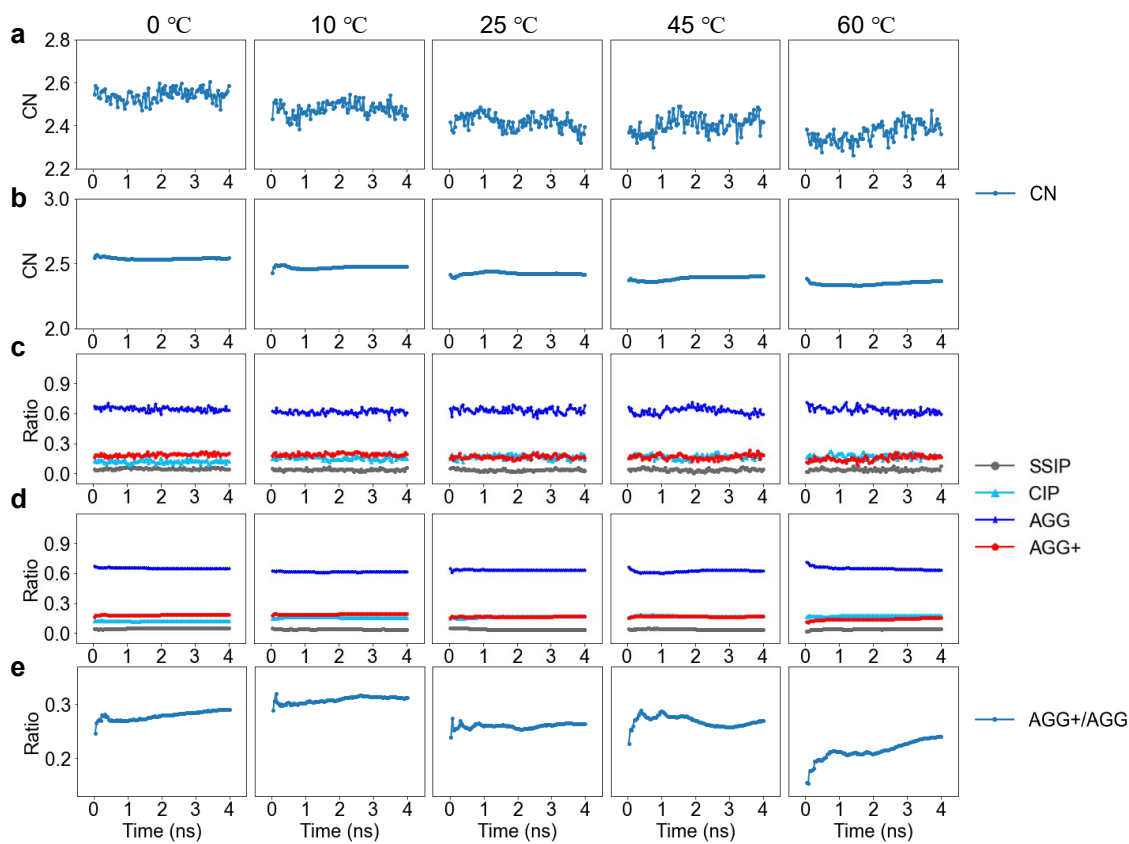

**Supplementary Fig. 24 | Statistics of coordination numbers (CN) and aggregation ratios for HCE (LiFSI-1.4DME)**. **a**, MD-calculated FSI$^-$ *vs* Li$^+$ CN as function of time during production runs (the last 4 ns simulations). **b**, Time-averaged FSI$^-$ *vs* Li$^+$ CN as function of time. **c**, MD-calculated aggregation ratios (SSIP, CIP, AGG and AGG+) as function of time. **d**, Time-averaged aggregation ratios (SSIP, CIP, AGG and AGG+) as functions of time. **e**, AGG+/AGG ratio as functions of time.

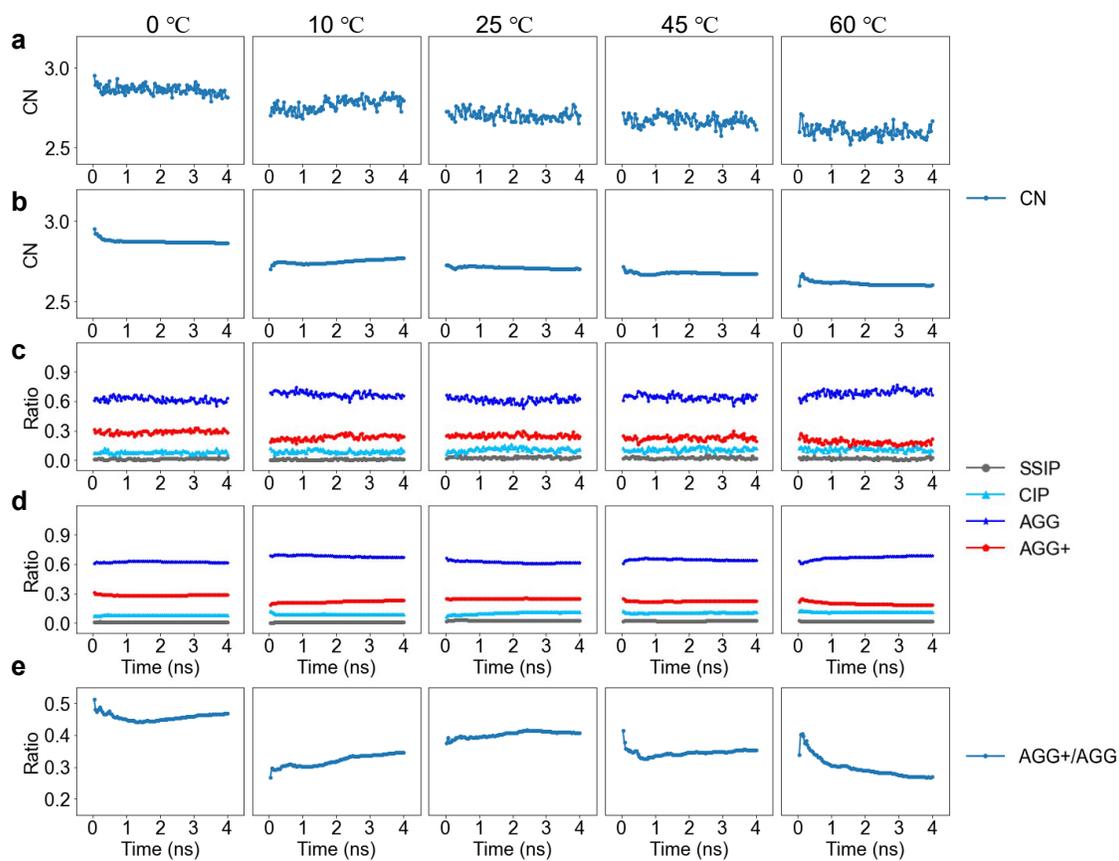

**Supplementary Fig. 25 | Statistics of coordination numbers (CN) and aggregation ratios for HCE (LiFSI-1.2DME)**. **a**, MD-calculated $FSI^-$ *vs* $Li^+$ CN as function of time during production runs (the last 4 ns simulations). **b**, Time-averaged $FSI^-$ *vs* $Li^+$ CN as function of time. **c**, MD-calculated aggregation ratios (SSIP, CIP, AGG and AGG+) as function of time. **d**, Time-averaged aggregation ratios (SSIP, CIP, AGG and AGG+) as function of time. **e**, AGG+/AGG ratio as function of time.

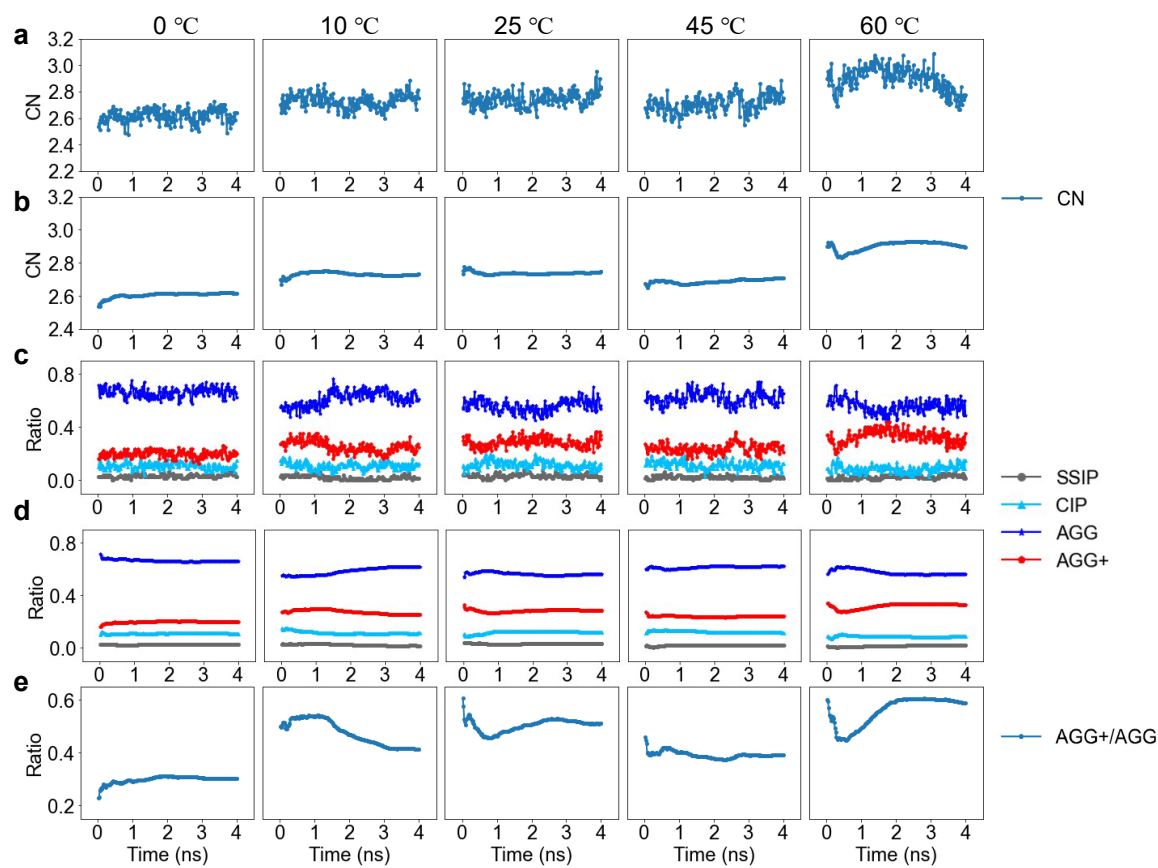

**Supplementary Fig. 26 | Statistics of coordination numbers (CN) and aggregation ratios for LHCE (LiFSI-1.2DME-2TFEO)**. **a**, MD-calculated $FSI^-$ vs $Li^+$ CN as function of time during production runs (the last 4 ns simulations). **b**, Time-averaged $FSI^-$ vs $Li^+$ CN as function of time. **c**, MD-calculated aggregation ratios (SSIP, CIP, AGG and AGG+) as function of time. **d**, Time-averaged aggregation ratios (SSIP, CIP, AGG and AGG+) as function of time. **e**, AGG+/AGG ratio as function of time.

## 4. Additional Convergence test for LHCE

The error bars are generally small in homogenous LCE and HCE and become larger in heterogenous LHCE. Thus, additional 10 ns NPT dynamics are run (Supplementary Fig. 25). The conclusion holds considering the error bars, including the occurrence of a peak value of AGG+/AGG at room temperature, which show larger scattering. This is likely due to the smaller cluster sizes in MD simulations compared to experiments.

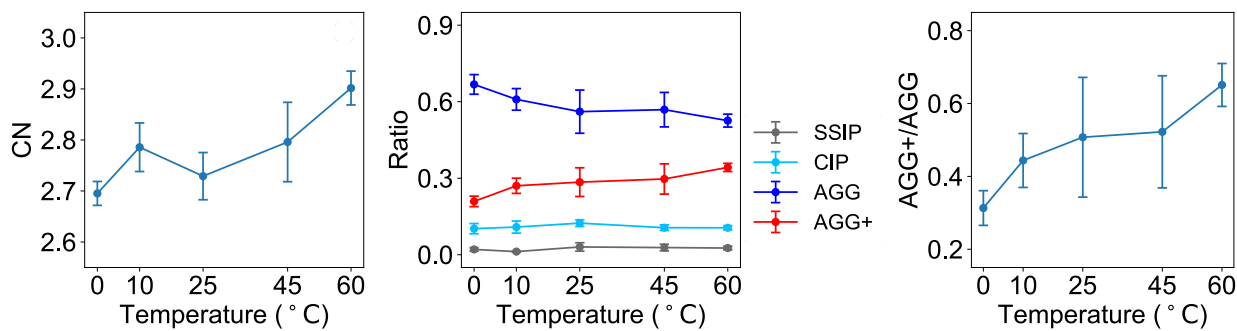

**Supplementary Fig. 27 | Time-averaged coordination number (CN) and aggregate types as a function of temperature for an extended 10.0 ns NPT simulation of LHCE (LiFSI-1.2DME-2TFEO). a**, Time-averaged FSI$^-$ *vs* Li$^+$ CN, **b**, ratios of aggregates (SSIP, CIP, AGG and AGG+), and **c**, AGG+/AGG ratios.

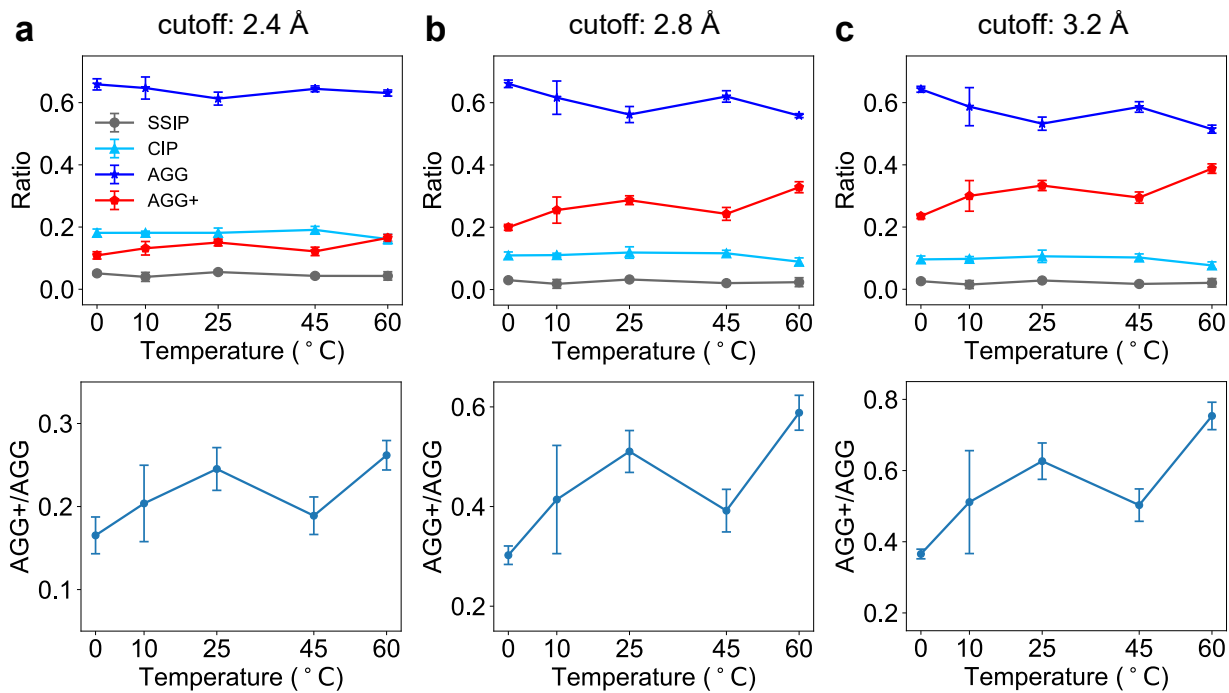

**Supplementary Fig. 28 | MD tests on different cutoffs used for calculating salt-solvent cluster ratios.** Calculated ratios (upper panels) of SSIP, CIP, AGG, and AGG+ as a function of temperature, as well as AGG+/AGG ratio (lower panels) as a function of temperature using the cutoff of **a**, 2.4 Å, **b**, 2.8 Å and **c**, 3.2 Å. It is seen that the trend does not change using different cutoffs (2.4 Å, 2.8 Å and 3.2 Å) when calculating the ratios of aggregates. The cutoff of 2.8 Å is used in this work. LHCE (LiFSI-1.2DME-2TFEO) was used to exemplify the chosen cutoff.

# References

1	Cao, X. *et al.* Optimization of fluorinated orthoformate based electrolytes for practical high-voltage lithium metal batteries. *Energy Storage Mater* **34**, 76-84, doi:10.1016/j.ensm.2020.08.035 (2021).
2	Su, L. S. *et al.* Uncovering the Solvation Structure of LiPF6-Based Localized Saturated Electrolytes and Their Effect on LiNiO2-Based Lithium-Metal Batteries. *Adv Energy Mater* **12**, doi:10.1002/aenm.202201911 (2022).
3	Adams, B. D., Zheng, J. M., Ren, X. D., Xu, W. & Zhang, J. G. Accurate Determination of Coulombic Efficiency for Lithium Metal Anodes and Lithium Metal Batteries. *Adv Energy Mater* **8**, doi:10.1002/aenm.201702097 (2018).
4	Piao, N. *et al.* Countersolvent Electrolytes for Lithium-Metal Batteries. *Adv Energy Mater* **10**, doi:10.1002/aenm.201903568 (2020).
5	Cao, X. *et al.* Effects of fluorinated solvents on electrolyte solvation structures and electrode/electrolyte interphases for lithium metal batteries. *P Natl Acad Sci USA* **118**, doi:10.1073/pnas.2020357118 (2021).
6	Cao, X. *et al.* Monolithic solid-electrolyte interphases formed in fluorinated orthoformate-based electrolytes minimize Li depletion and pulverization. *Nat Energy* **4**, 796-805, doi:10.1038/s41560-019-0464-5 (2019).
7	Ren, X. D. *et al.* Enabling High-Voltage Lithium-Metal Batteries under Practical Conditions. *Joule* **3**, 1662-1676, doi:10.1016/j.joule.2019.05.006 (2019).
8	Hou, Z. P. *et al.* Protecting Li Metal Anode While Suppressing "Shuttle Effect" of Li-S Battery Through Localized High-Concentration Electrolyte. *J Electron Mater* **51**, 4772-4779, doi:10.1007/s11664-022-09751-z (2022).
9	Chen, S. R. *et al.* High-Voltage Lithium-Metal Batteries Enabled by Localized High-Concentration Electrolytes. *Adv Mater* **30**, doi:10.1002/adma.201706102 (2018).
10	Liu, T. *et al.* Low-Density Fluorinated Silane Solvent Enhancing Deep Cycle Lithium-Sulfur Batteries' Lifetime. *Adv Mater* **33**, doi:10.1002/adma.202102034 (2021).
11	Zheng, J. *et al.* High-Fluorinated Electrolytes for Li-S Batteries. *Adv Energy Mater* **9**, doi:10.1002/aenm.201803774 (2019).
12	Tan, S. *et al.* Isoxazole-Based Electrolytes for Lithium Metal Protection and Lithium-Sulfurized Polyacrylonitrile (SPAN) Battery Operating at Low Temperature. *J Electrochem Soc* **169**, doi:10.1149/1945-7111/ac58c5 (2022).
13	Huangzhang, E. *et al.* A localized high-concentration electrolyte with lithium bis(fluorosulfonyl) imide (LiFSI) salt and F-containing cosolvents to enhance the performance of Li||LiNi0.8Co0.1Mn0.1O2 lithium metal batteries. *Chem Eng J* **439**, doi:10.1016/j.cej.2022.135534 (2022).
14	Chen, S. R. *et al.* High-Efficiency Lithium Metal Batteries with Fire-Retardant Electrolytes. *Joule* **2**, 1548-1558, doi:10.1016/j.joule.2018.05.002 (2018).
15	Ren, X. D. *et al.* Localized High-Concentration Sulfone Electrolytes for High-Efficiency Lithium-Metal Batteries. *Chem-Us* **4**, 1877-1892, doi:10.1016/j.chempr.2018.05.002 (2018).
16	Borodin, O. & Smith, G. D. Quantum Chemistry and Molecular Dynamics Simulation Study of Dimethyl Carbonate: Ethylene Carbonate Electrolytes Doped with LiPF6. *J Phys Chem B* **113**, 1763-1776, doi:10.1021/jp809614h (2009).

17	Qian, J. F. *et al.* High rate and stable cycling of lithium metal anode. *Nat Commun* **6**, doi:10.1038/ncomms7362 (2015).
18	Hayashi, K., Nemoto, Y., Tobishima, S. & Yamaki, J. Electrolyte for high voltage Li/LiMn1.9Co0.1O4 cells. *J Power Sources* **68**, 316-319, doi:10.1016/S0378-7753(97)02636-0 (1997).